\documentclass[11pt,a4paper,reqno]{amsart}

\usepackage[latin1]{inputenc}
\usepackage[english]{babel}
\usepackage{amsmath}
\usepackage[numbers]{natbib}
\usepackage{graphicx}
\usepackage{amsmath,bm}
\usepackage{booktabs}
\usepackage{graphicx}
\usepackage{multirow}
\usepackage{setspace}
\usepackage{braket}
\usepackage[hmargin=3cm,vmargin={3cm,4cm}]{geometry}

\usepackage{etoolbox}
\newcommand*{\affaddr}[1]{#1} 
\newcommand*{\affmark}[1][*]{\textsuperscript{#1}}

\begin{document}

\title{Multiscale modelling of oxygenic photogranules}

\author[A.~Tenore et al.]{A.~Tenore\protect\affmark[1] \and M.R.~Mattei\affmark[1] \and L.~Frunzo\affmark[1]}

 \maketitle

 {\footnotesize
  \begin{center}
\affaddr{\affmark[1]Department of Mathematics and Applications, University of Naples "Federico II", via Cintia 1, Montesantangelo, 80126, Naples, Italy}\\
Corrisponding authors: A.~Tenore, \texttt{alberto.tenore@unina.it}; M.R.~Mattei,
          \texttt{mariarosaria.mattei@unina.it}
 \end{center}}

\begin{abstract}

This work presents a mathematical model which describes both the genesis and growth of oxygenic photogranules (OPGs) and the related treatment process. The photogranule has been modelled as a free boundary domain with radial symmetry, which evolves over time as a result of microbial growth, attachment and detachment processes. A system of hyperbolic and parabolic partial differential equations (PDEs) have been considered to model the advective transport and growth of sessile biomass and the diffusive transport and conversion of soluble substrates. The reactor has been modelled as a sequencing batch reactor (SBR), through a system of first order impulsive ordinary differential equations (IDEs). Phototrophic biomass has been considered for the first time in granular biofilms, and cyanobacteria and microalgae are taken into account separately, to model their differences in growth rate and light harvesting and utilization. To describe the key role of cyanobacteria in the formation process of photogranules, the attachment velocity of all suspended microbial species has been modelled as a function of the cyanobacteria concentration in suspended form. The model takes into account the main biological aspects and processes involved in OBGs based systems: heterotrophic and photoautotrophic activities of cyanobacteria and microalgae, metabolic activity of heterotrophic and nitrifying bacteria, microbial decay, EPS secrection, diffusion and conversion of soluble substrates (inorganic and organic carbon, ammonia, nitrate and oxygen), symbiotic and competitive interactions between the different microbial species, day-night cycle, light diffusion and attenuation across the granular biofilm and photoinhibion phenomena. The model has been integrated numerically, investigating the evolution and microbial composition of photogranules and the treatment efficiency of the OPGs-based system. The results  show the consistency of the model and confirm the purifying effectiveness of the OPGs technology, by analyzing the effects of the wastewater influent composition and light conditions on the process.
 
\end{abstract}

\maketitle

\section{Introduction} \label{n5.1}
 \
Nowadays, great interest is addressed to innovative biological systems in the field of wastewater treatment, accounted to improve treatment efficiencies and, at the same time, reduce the operating and management costs of traditional systems. In this context, several studies have been carried out to investigate treatment potentialities of granular biofilms \cite{trego2019granular,nicolella2000wastewater}. Biofilm granules are described as spherical, dense aggregate composed by microbial organisms embedded in a self-produced matrix of extracellular polymeric substances (EPS) \cite{trego2020granular}. The process leading to the granule formation is known as \textit{de novo} granulation and consists in the aggregation of microbial cells and flocs under particular conditions. Compared to conventional biomass flocs, biofilm granules have higher densities, which lead to microscale gradients of the substrates concentration and, consequently, to the formation of different microbial niches along the radius \cite{weissbrodt2013assessment}. As a result, biofilm granules are extremely heterogeneous ecosystems populated by a broad diversity of microbial species. Specifically, the complexity, variability and multiplicity of microbial metabolic activities lead to a dense and intricate network of symbiotic and competitive interactions between the microbial species inhabiting the granule ecosystem.

For several years, full-scale biological systems based on aerobic, anaerobic and anammox granular biofilms have been developed \cite{seghezzo1998review,pronk2015full,van2007startup}, while only more recently, attention has been paid to oxygenic photogranules (OPGs) \cite{milferstedt2017importance,stauch2017role}, biofilm granules developed in presence of light and constituted by a relevant phototrophic component. OPGs-based systems show huge advantages compared to conventional systems. Above all, they do not require aeration and this leads to lower energy costs compared to all aerobic systems. Indeed, the phototrophic production of $O_2$ meets the demand of heterotrophic and nitrifying bacteria and allows the oxidation of carbon and nitrogen compounds without the need of external $O_2$ sources \cite{milferstedt2017importance}. Due to their high density, photogranules have settling properties higher than suspended biomass and this leads to a more efficient separation of the biomass from the treated water \cite{abouhend2018oxygenic}, and allows to reach higher biomass concentrations without the risk of re-suspension \cite{baeten2019modelling}. Specifically, OPGs-based systems are more convenient compared to suspended systems, whose main drawback is represented by the complex and expensive separation phase, conventionally based on harvesting procedures \cite{roeselers2008phototrophic}. Moreover, OPGs-based systems involve sessile biomass arranged in spherical and constantly moving granules and this mitigates boundary layer resistances and enhances the mass transfer of substrates across the biofilm photogranule \cite{baeten2019modelling}. In addition, photogranules biomass deriving from wastewater treatments represents a renewable energy source to be used in industrial applications such as the production of biofuels, chemicals, and nutraceuticals \cite{milferstedt2017importance}.

The granulation process is complex and not fully understood. Over time, several theories have been proposed to explain the formation process of aerobic, anaerobic and anammox biofilm granules \cite{pol2004anaerobic,beun1999aerobic}. In this context, hydrodynamic conditions are regarded as the key factor of the process \cite{liu2002essential}, although some works assign a decisive role to specific microbial species \cite{pol2004anaerobic}. In the case of oxygenic photogranules, some experimental evidences have highlighted their formation in batch reactors, under hydrostatic conditions \cite{milferstedt2017importance,ansari2019effects}. In this regard, the discriminating factor in the photogranulation process appears to be the presence of cyanobacteria \cite{milferstedt2017importance,abouhend2018oxygenic,abouhend2019growth}. Due to their motility and filamentous structure, cyanobacteria are able to arrange in mat-like layers which encloses other microbial species in rigid, spherical structures \cite{ansari2019effects}. Some studies have reported high chance of success of the photogranulation process in the case of inocula rich in cyanobacteria, which reduces drastically in the case of inocula poor in cyanobacteria \cite{hann2018factors}.

Although full-scale OPGs-based systems have not yet been developed, many experimental studies based on lab scale systems have been carried out to investigate the main aspects of the OPGs formation and their treatment potential \cite{milferstedt2017importance,ansari2019effects,stauch2017role,downes2019success,ahmad2017stability}. As well as aerobic granules, photogranules are cultivated in sequencing batch reactors (SBRs). The experimental results recommend this system as one of the most promising technologies in the field of wastewater treatment \cite{abouhend2018oxygenic,hann2018factors}. 

Several models have been proposed to describe the dynamics of phototrophic biofilms growing on solid supports and investigate the main mechanisms which drive the process \cite{clarelli2013fluid,clarelli2016fluid,li2016investigating,polizzi2017time,wolf2007kinetic,munoz2014modeling}, while others models have focused on the description of aerobic, anaerobic or anammox biofilm granules \cite{de2007kinetic,odriozola2016modeling,volcke2010effect}, by considering a spherical geometry with radial symmetry. Nevertheless, none of the existing models addressed the dynamics of OPGs. In this context, the present work introduces a mathematical model which describes for the first time the formation and evolution of OPGs, and the related treatment process. 

The model is based on a multiscale approach \cite{mavsic2014modeling} and incorporates the mesoscopic granular biofilm model with the macroscopic reactor mass balances. The granular biofilm model is formulated as a free boundary problem applied to a spherical domain with radial symmetry \cite{wanner1986multispecies,tenore2021multiscale}. As introduced in \cite{d2019free}, a vanishing value is adopted for the initial granule dimension in order to describe the evolution of the granular biofilm from its genesis. In addition, the substrate dynamics within the SBR are modelled through a system of first order impulsive ordinary differential equations (IDEs) \cite{ferrentino2018process}.

To correctly model oxygenic photogranules, it is necessary to differentiate cyanobacteria from microalgae. Models that take into account the growth of phototrophic biomass in engineering systems include all phototrophic species under a common model component, and suppose the entire phototrophic community to have the same metabolic activity. Indeed, the different phototrophic groups have similar metabolic functions and the engineering purposes do not require this distinction. Nevertheless, cyanobacteria play a key role in the process of formation and growth of photogranules, due to characteristics and abilities that microalgae lack. Therefore, cyanobacteria and microalgae are modelled as two distinct components, taking into account the main differences in metabolic activity \cite{tilzer1987light}, attachment properties \cite{milferstedt2017importance} and light harvesting \cite{abouhend2018oxygenic,abouhend2019growth,ting2002cyanobacterial}. Specifically, as mentioned above, it is realistically assumed that cyanobacteria govern the granulation process due to their attachment properties and promote also the presence of other species within the granules. Consequently, the attachment velocities of all suspended microbial species are not assumed to be constant, as in \cite{tenore2021multiscale}, but functions of the concentration of suspended cyanobacteria present in the system.

Light influences the metabolic activity of cyanobacteria and microalgae and is included as a model variable. The light intensity is modelled as a piecewise-constant time function within the reactor, to simulate the day-night cycle, and as a function of time and space within the granule, by considering attenuation phenomena through the Lambert-Beer law. The metabolic activity of cyanobacteria and microalgae has been modelled by taking into account the main processes and factors: photoautotrophic growth with consumption of ammonium or nitrate under light conditions \cite{wolf2007kinetic}, heterotrophic pathway in dark conditions \cite{wagner2016towards}, release of dissolved organic matter \cite{el2017photorespiration}, photoinhibition phenomena \cite{steele1962environmental}, inhibition of oxygen on the photosynthetic activity \cite{li2016investigating}. In addition to cyanobacteria and microalgae, other microbial components have been assumed to compose the biofilm granule: aerobic and denitrifying heterotrophic bacteria, nitrifying bacteria, EPS and inert material. Moreover, the diffusion and conversion of the following soluble substrates are considered: inorganic carbon ($IC$), organic carbon ($DOC$), nitrate ($NO_3$), ammonia ($NH_3$), oxygen ($O_2$).  The main microbial interaction involving the mentioned microbial species and dissolved substrates and taking place in the photogranule ecosystem are discussed in the following sections.

 The model has been integrated numerically, to investigate the main biological aspects of OPGs evolutions and the performances of OPGs-based systems, under different influent compositions and light conditions. The work is organized as follows. The mathematical model is introduced and described in Section \ref{n5.2}. Numerical studies and results are reported in Section \ref{n5.3} and discussed in Section \ref{n5.4}. All reaction terms and the related stoichiometric coefficients are reported in Appendix A.


 \section{Mathematical Model} \label{n5.2}
\

The model describes the evolutionary process of oxygenic photogranules in an SBR reactor, including their formation, induced by the granulation of the suspended biomass present in the reactor. The OPGs based system has been described by modelling both the mesoscopic granules and the macroscopic SBR reactor and by considering how they influence each other through biomass and substrate exchange fluxes. The two submodels that describe the dynamics within the granular biofilm and the SBR reactor are reported in the following.

 \subsection{Mesoscopic Granular Biofilm model} \label{n5.2.1} 
 \
The model introduced in \cite{tenore2021multiscale} is used to describe the growth of the granular biofilm, maintaining the same equations and assumptions. It is assumed that all granules are subject to same phenomena, therefore, at any time it is possible to identify in the system $N_G$ granules having same properties (density, dimension, microbial composition etc.). Specifically, each granule has been modelled as a spherical free boundary domain with radial symmetry, which evolves due to the growth of sessile biomass, the attachment flux from bulk liquid to granule $\sigma_{a}$ and detachment flux from granule to bulk liquid $\sigma_{d}$. Thus, the spatial variability of the quantities involved can be fully described by the radial coordinate $r$, locating the granule centre at $r = 0$.

The components considered within the biofilm granule, expressed in terms of concentration, are the sessile biomasses constituting the solid matrix $X_i(r,t)$ and the soluble substrates $S_j(r,t)$. All sessile microbial species are supposed to have the same constant density $\rho$ and the volume fraction of each individual species is achieved by dividing its concentration $X_i$ by $\rho$. Notably, the fractions of volume occupied by biomass are constrained to add up to unity, $\sum_{i = 1}^{n} f_i = 1$ \cite{rahman2015mixed}.

A system of hyperbolic PDEs describes the transport and growth of sessile microbial species within the granule:

   \[	 
\frac{\partial f_i(r,t)}{\partial t} +  u(r,t)\frac{\partial f_i(r,t)}{\partial r}
			 =r_{M,i}(r,t,{\bf f},{\bf S}) - f_i(r,t) \sum_{i=1}^{n}r_{M,i}(r,t,\textbf{f},\textbf{S}),
  \]
\begin{equation}                                        \label{5.2.1}
		 i=1,...,n, 0 \leq r \leq R(t),\ t>0,
\end{equation}	

  \begin{equation}                                        \label{5.2.2}  
   f_i(R(t),t) = \frac{v_{a,i}\psi^*_i(t)}{\sum_{i=1}^{n}v_{a,i}\psi^*_i(t)}, \ i=1,...,n,\ t>0, \ \sigma_{a}-\sigma_{d}>0,
  \end{equation} 

where $r_{M,i}(r,t,{\bf f},{\bf S})$ is the growth rate of the $i^{th}$ sessile biomass, $v_{a,i}$ is the attachment velocity and $\psi^*_i(t)$ is the concentration within the bulk liquid of the $i^{th}$ suspended biomass, while $u(r,t)$ is the biomass velocity and is governed by the following equation:
    
    \begin{equation}                                        \label{5.2.3}		
	     \frac{\partial u(r,t)}{\partial r} = -\frac{2 u(r,t)}{r} + \sum_{i=1}^{n}r_{M,i}(r,t,\textbf{f},\textbf{S}), \ 0 < r < R(t),\ t>0,
	\end{equation}
	
    \begin{equation}                                        \label{5.2.4}		
	     u(0,t)=0, \ t>0.
	\end{equation}
  
 Eq. \ref{5.2.2} represents the boundary condition for Eq. \ref{5.2.1} at the interface granule-bulk liquid and holds only when attachment flux is higher than detachment flux. Indeed, 
 in the opposite case, the biomass concentration at the interface is regulated exclusively by the internal points of the biofilm domain. The evolution of the free boundary domain is described by the radius of the biofilm granule $R(t)$ which varies according to the following equation, derived from the global mass balance on the granule volume:

    \begin{equation}                                       \label{5.2.5}
  \dot R(t)
    = \sigma_a(t)-\sigma_d(t) + u(R(t),t),
 \end{equation}
 
    \begin{equation}                                       \label{5.2.6}
  R(0)= 0.
 \end{equation}

 All biomass is initially supposed in suspended form by setting a vanishing value as initial condition of $R(t)$. The granulation process is initiated by the attachment flux of suspended biomass, assumed linearly dependent on the concentration of suspended biomasses within the bulk liquid $\psi^*_i(t)$:

  \begin{equation}                                      \label{5.2.7}
    \sigma_a(t)=\sum_{i=1}^{n} \sigma_{a,i}(t)=\frac{\sum_{i=1}^{n}v_{a,i}\psi^*_i(t)}{\rho}. 
 \end{equation}
 
Meanwhile, the detachment flux is modelled as \cite{abbas2012longtime}

\begin{equation}                                        \label{5.2.8}
	    \sigma_d(t)=\lambda R^2(t),
\end{equation} 

where $\lambda$ is the constant detachment coefficient.

The soluble substrates diffusion and conversion within the granule are governed by the following system of parabolic PDEs:

   \[	 
  \frac{\partial S_j(r,t)}{\partial t}-D_{S,j}\frac{\partial^2 S_j(r,t)}{\partial r^2} - \frac{2 D_{S,j}}{r} \frac{\partial S_j(r,t)}{\partial r}=
   r_{S,j}(r,t,{\bf f},{\bf S}),
  \]
 \begin{equation}                                        \label{5.2.9}
	   \ j=1,...,m, 0 < r < R(t),\ t>0,
 \end{equation}

 \begin{equation}                                        \label{5.2.10}
   \frac{\partial S_j}{\partial r}(0,t)=0,\ S_j(R(t),t))=S^*_j(t),\ j=1,...,m,\ t>0,
 \end{equation}

where $D_{S,j}$ is the diffusion coefficient in biofilm, $r_{S,j}(r,t,{\bf f},{\bf S})$  is the conversion rate and $S^*_j(t)$ is the concentration of substrate $j$ in the bulk liquid.

 \subsection{Macroscopic SBR model} \label{n5.2.2} 
 \
The second submodel has been developed to describe the dynamics of the sequencing batch reactor (SBR) where the growth of $N_G$ identical photogranules occurs. SBR reactors are based on a cyclic operation, in which the wastewater influent is fed and treated discontinuously: in the first phase (filling) a volume of wastewater is fed within the reactor; in the second phase (reaction), such volume is biologically treated by means of the biomass present in the system; the biomass-liquid separation takes place in the third phase (settling); lastly, in the last phase (emptying) the purified supernatant is removed and the reactor is refilled with a new wastewater volume to be treated.  The model components considered in the bulk liquid are the concentrations of suspended biomasses $\psi^*_i(t)$ and the concentrations of soluble substrates $S^*_j(t)$.

To model this system some assumptions have been introduced:

\begin{itemize}
 \item The filling, settling and emptying phases take place instantaneously and influence the process through an instantaneous change of state of the system, therefore the cycle time corresponds to the reaction time.
 \item The reactor is completely mixed, hence, the variables involved are functions of time and not of space.
 \item The volume of granular and suspended biomass is neglected (the reactor volume is equal to the liquid volume).
 \item All sessile biomass remains in the reactor during the emptying phase (100\% settling efficiency of biofilm granules); suspended biomass has a partial settling efficiency.
 \item During the emptying phase the reactor is only partially emptied (emptying/refilling ratio less than 1).
 \end{itemize}
  
 Mathematically, the SBR configuration is modelled through a system of first order impulsive ordinary differential equations (IDEs) \cite{ferrentino2018process}. Each impulsive differential equation is based on three components: the continuous-time differential equation which governs the state of the system between impulses; the impulse equation, which describes an impulsive jump and is defined by a jump function at the instant the impulse occurs; and the jump criterion, which defines a set of jump events in which the impulse equation is active \cite{ferrentino2018process}. These equations have been applied to $\psi^*_i(t)$ and $S^*_j(t)$:
 
    \[	 
	   V \dot \psi^*_i(t)=-\sigma_{a,i}(t) \rho A(t) N_G + r^*_{\psi,i}(t,{\bm \psi^*},{\bf S^*}),\
  \]	
 \begin{equation}                                       \label{5.2.11}
 t \in [0,T], \ t \neq t_k, \ \psi^*_i(0)=\psi^*_{i,0}, \ i=1,...,n,
 \end{equation}
 
    \[	 
	   V \dot S^*_j(t)= - A(t) N_G D_{S,j} \frac{\partial S_j(R(t),t)}{\partial r} +r^*_{S,j}(t,{\bm \psi^*},{\bf S^*}),
  \]	
 \begin{equation}                                       \label{5.2.12}
t \in [0,T], \ t \neq t_k, \ S^*_j(0)=S^*_{j,0}, \ j=1,...,m,
 \end{equation}

 \begin{equation}                                       \label{5.2.13}
\Delta \psi^*_i(t_k)=\psi^*_i(t^+_k)-\psi^*_i(t^-_k)= - \gamma \psi^*_i(t^-_k), \ k=1,...,h, \ i=1,...,n,
 \end{equation}
 
  \begin{equation}                                       \label{5.2.14}
\Delta S^*_j(t_k)=S^*_j(t^+_k)-S^*_j(t^-_k)= - \omega S^*_j(t^-_k) + \omega S^{in}_j, \ k=1,...,h, \ j=1,...,m,
 \end{equation}
 
 where $r^*_{\psi,i}(t,{\bm \psi^*},{\bf S^*})$ and  $r^*_{S,j}(t,{\bm \psi^*},{\bf S^*})$ are the conversion rates for $\psi_i^*$ and $S_j^*$, respectively; $V$ is the reactor volume; $A(t)$ is the area of the spherical granule and is equal to $4 \pi R^2(t)$; $\psi^*_{i,0}$ and $S^*_{j,0}$ are the initial concentrations of the $i^{th}$ suspended species and the $j^{th}$ soluble substrate within the bulk liquid, respectively; $S^{in}_j$ is the concentration of the $j^{th}$ substrate in the influent; $\gamma$ is the fraction of suspended biomass removed during the emptying phase; $\omega$ is the emptying/refilling ratio;  $0=t_0 < t_1 < t_2 < ... < t_h < t_{h+1}=T$, $t_{k+1}-t_{k}=\tau$; $\tau$ is the duration of a cycle; $\psi^*_i(t^+_k)$, $S^*_j(t^+_k)$, $\psi^*_i(t^-_k)$, $S^*_j(t^-_k)$ are the right and left limits of $\psi^*_i$ and $S^*_j$ at time $t_k$.

 \subsection{Model components} \label{n5.2.3} 
 \
As previously described, the formation process of oxygenic photogranules is strongly influenced by the presence of cyanobacteria in the inoculum and this requires to define two distinct model components, the cyanobacteria and the microalgae. The components supposed to constitute the solid matrix of the biofilm are (expressed in terms of concentrations): cyanobacteria $X_{C}(r,t)$, microalgae $X_{A}(r,t)$, heterotrophic bacteria $X_{H}(r,t)$, nitrifying bacteria $X_{N}(r,t)$, EPS $X_{EPS}(r,t)$ and inert material $X_{I}(r,t)$. The soluble substrates considered are (expressed in terms of concentrations): inorganic carbon $S_{IC}(r,t)$, organic carbon $S_{DOC}(r,t)$, nitrate $S_{NO_3}(r,t)$, ammonium $S_{NH_4}(r,t)$ and dissolved oxygen $S_{O_2}(r,t)$. The same soluble substrates are considered in the bulk liquid and their concentration is expressed as $S^*_j(t)$, $j \in \{IC, DOC, NO_3, NH_4, O_2\}$. The suspended microbial species supposed to populate the bulk liquid are (expressed in terms of concentrations): cyanobacteria $\psi^*_{C}(t)$, microalgae $\psi^*_{A}(t)$, heterotrophic bacteria $\psi^*_{H}(t)$ and nitrifying bacteria $\psi^*_{N}(t)$. Inert formation and EPS production are processes occuring also within the bulk liquid. Indeed, beyond the key role in biofilm growth, EPS contributes to the formation of suspended flocs \cite{laspidou2002unified}. In any case, it is likely to assume that EPS production by suspended biomass is much lower than sessile production \cite{laspidou2002unified} and it has been neglected. Furthermore, inert material in suspended form does not play any role on the evolution of granules and on system performances, therefore it has been not included in the model. Heterotrophic bacteria include aerobic heterotrophs which grow by consuming oxygen and denitrifying heterotrophs which grow by consuming nitrate and are inhibited by oxygen. Since cyanobacteria and microalgae need light to develop their metabolic activity, light intensity $I(r,t)$ has been included as a model variable. $I$ is assumed to be constant and equal at every point of the bulk liquid, while it attenuates within the granules according to the Lambert-Beer law:

\begin{equation}                                        \label{5.2.15}
I(r,t)=I_0 \ e^{-k_{tot}(R(t)-r)\rho}, \ 0 \leq r \leq R(t),\ t>0,
\end{equation}

where $I_0$ is the fixed light intensity at the granule surface and $k_{tot}$ is the light attenuation coefficient. 

 \subsection{Modelling attachment} \label{n5.2.4} 
 \
The photogranulation process is governed by cyanobacteria in suspended form which aggregate due to the motility, the filamentous structure and their ability to secrete EPS. During this process, cyanobacteria envelop other microbial species leading to granules populated by an heterogeneous microbial community \cite{ansari2019effects}. This process is modelled in a deterministic way: the cyanobacteria attachment flux is assumed proportional to the concentration and the attachment velocity of suspended cyanobacteria ($\psi^*_{C}$ and $v_{a,C}$, respectively):

  \begin{equation}                                      \label{5.2.16}
    \sigma_{a,C}(t)=v_{a,C} \frac{\psi^*_{C}(t)}{\rho}. 
 \end{equation}
 
With regard to the attachment flux of other species, a new expression is here introduced to take into account the role of cyanobacteria. For this reason, the attachment flux of microalgae, heterotrophs and nitrifiers takes the following form:

  \begin{equation}                                      \label{5.2.17}
    \sigma_{a,i}(t)=v_{a,i}(\psi^*_{C}(t)) \frac{\psi^*_{i}(t)}{\rho}, \ i \in \{A, H, N \}, 
 \end{equation}
 
  \begin{equation}                                      \label{5.2.18}
    v_{a,i}(\psi^*_{C}(t))= \frac{v^0_{a,i} \psi^*_{C}(t)}{K_{C}+\psi^*_{C}(t)}, \ i \in \{A, H, N \},
  \end{equation}

where $v^0_{a,i}$ is the maximum attachment velocity of the $i^{th}$ suspended species and $K_{C}$ is the cyanobacteria half saturation constant on the attachment of microalgae, heterotrophs and nitrifiers. 


 \subsection{Microbial kinetics of cyanobacteria and microalgae} \label{n5.2.5} 
 \
In general, the metabolic activity of microalgae and cyanobacteria is extremely complex and species-specific and can be characterized by photoautotrophic, heterotrophic and mixotrophic pathways \cite{roostaei2018mixotrophic}, depending on a multitude of factors, such as taxonomy, light conditions, alternation of light and dark, temperature, availability of nutrients \cite{borowitzka2016physiology}. For the engineering and ecological purposes of this work, the main biological processes have been modelled. In presence of light, the metabolism of microalgae and cyanobacteria is supposed photoautotrophic and based on photosynthetic activity: using light energy, cyanobacteria and microalgae grow by consuming $IC$ and $NH_3$, release $DOC$ and $O_2$ and secrete EPS. In lack or shortage of $NH_3$, cyanobacteria and microalgae are supposed to grow by using $NO_3$ as a nitrogen source \cite{wolf2007kinetic}. The model takes into account the inhibition induced by the presence of $O_2$ on photosynthetic activity. In absence of light, cyanobacteria and microalgae are supposed to have a heterotrophic metabolic activity: they grow by consuming $O_2$, $DOC$ and $NH_3$ and produce $IC$. 


In addition to the different granulation ability, some differences in the metabolic kinetics of cyanobacteria and microalgae has been considered, based on literature. As reported in \cite{rossi2015role}, although microalgae and other microbial species contribute to the EPS production, cyanobacteria are seen as the main EPS contributors throughout the biofilm development. The model takes into account this evidence by considering a higher production of EPS by cyanobacteria compared to microalgae and other species. In accordance with \cite{tilzer1987light}, the maximum growth rate of cyanobacteria has been assumed lower than microalgae. Furthermore, differences in the harvesting and utilization of light are widely documented in the literature \cite{abouhend2018oxygenic,abouhend2019growth,ting2002cyanobacterial}: the optimal light intensity for the photoautotrophic metabolic activity of cyanobacteria is lower than microalgae and their adaptability to extreme light conditions appears to be enhanced. These differences are take into account by setting the parameters of the light dependency coefficient \cite{steele1962environmental} based on the phototrophic microbial species and by introducing the additional parameter $\eta_i$ in the original formulation, which represents adaptability to non-optimal light conditions:

\begin{equation}                                        \label{5.2.19}
\phi_{I,i}(r,t) = (\frac{I(r,t)}{I_{opt,i}})^{\eta_i} \ e^{(1-(\frac{I(r,t)}{I_{opt,i}})^{\eta_i})}, \ i \in \{C,A \}, \ 0 \leq r \leq R(t),\ t>0,
\end{equation}

where $I_{opt,i}$ is the optimum light intensity and $\eta_i$ is the coefficient of adaptability to light. Specifically, $\eta_C < \eta_{A}=1$.

This formulation of the light dependence coefficient has been chosen because it also takes into account the phenomenon of photoinhibition, which limits photoautotrophic growth under adversely high light conditions.

 \subsection{Metabolic microbial interactions} \label{n5.2.6}
 \
The active biomasses (cyanobacteria, microalgae, heterotrophs and nitrifiers) grow by converting the soluble substrates, decay turning into inert material and secrete EPS. Due to their metabolic demand, biomasses cooperate and/or compete with each other. The most relevant interactions have been included in the model. As previously mentioned, the metabolic activity of cyanobacteria and microalgae is affected by light. Consequently, their interactions with the other species constituting the microbial community of the granule change according to the light conditions. In presence of light, cyanobacteria and microalgae promote the growth of nitrifiers and aerobic heterotrophs by producing $O_2$ and the growth of aerobic and anoxic heterotrophs by releasing $DOC$, while compete with nitrifiers for $IC$. Moreover, in absence or shortage of $NH_3$, they grow on $NO_3$, competing with anoxic heterotrophs. Under dark conditions, cyanobacteria and microalgae produce $IC$ for nitrifiers, while compete with aerobic heterotrophs and nitrifiers for $O_2$ and with all heterotrophs for $DOC$. In return, heterotrophic bacteria produce $IC$ necessary for the metabolism of cyanobacteria, microalgae and nitrifiers. Lastly, the latter produce $NO_3$ necessary for anoxic heterotrophs and for cyanobacteria and microalgae (in lack or shortage of reduced nitrogen). 

All the biological processes are included in the model through the reaction terms of Eqs. \ref{5.2.1}, \ref{5.2.3}, \ref{5.2.9}, \ref{5.2.11} and \ref{5.2.12}. The expressions of such terms are reported in Appendix A.

 
 \section{Numerical studies} \label{n5.3}
 \ 
  The model has been coded and implemented in MatLab, and integrated through numerical methods. Specifically, the method of characteristics has been used to track the biofilm expansion, a finite difference approximation has been adopted for the diffusion-reaction PDEs and a first-order approximation is used for the reactor impulsive equations. The time to compute the values of the unknown variables is in the order of hours to days, depending on the specific target simulation time $T$. 
 
 The numerical results on both the photogranule and the reactor scales have been analyzed. In this regard, three different numerical studies have been carried out and discussed in the following paragraphs. The first study has been focused on the treatment of a typical municipal wastewater and both the microbial characteristics of the biofilm granules and the treatment process have been investigated. In the second study, different types of influent wastewaters have been considered, to study how the influent composition affect the microbial abundance and distribution within the granules and to check the purifying efficiency of the system. Finally, the third study investigates the process under different light conditions and how these govern the microbial growth, with particular attention to the dualism between microalgae and cyanobacteria.
 
  \begin{table}[ht]
\begin{scriptsize}
\begin{spacing}{1.2}
 \begin{center}
 \begin{tabular}{llccc}
 \hline
{\textbf{Parameter}} & {\textbf{Definition}} & {\textbf{Unit}} & {\textbf{Value}} & {\textbf{Ref}}
 \\
 \hline
 $q_{max,A}$ & Maximum specific $O_2$ production rate by $A$ &  $kmol(O_2) kgCOD^{-1} d^{-1}$ & $0.074$ & \cite{wolf2007kinetic} \\
  $q_{max,C}$ & Maximum specific $O_2$ production rate by $C$ &  $kmol(O_2) kgCOD^{-1} d^{-1}$  & $0.037$  & (a) \\
  $\mu_{max,H}$ & Maximum specific growth rate for $H$ &  $d^{-1}$ & $4.8$  & \cite{munoz2014modeling}\\
  $\mu_{max,N}$ & Maximum specific growth rate for $N$ &  $d^{-1}$  & $1$ & \cite{munoz2014modeling} \\
 $k_{d,A}$ & Decay-inactivation rate for $A$  &  $d^{-1}$ & $0.1$  & \cite{wolf2007kinetic} \\
 $k_{d,C}$ & Decay-inactivation rate for $C$       &  $d^{-1}$ & $0.1$  & \cite{wolf2007kinetic} \\
 $k_{d,H}$  & Decay-inactivation rate for $H$ &  $d^{-1}$ & $0.1$  & \cite{wolf2007kinetic} \\
 $k_{d,N}$ & Decay-inactivation rate for $N$       &  $d^{-1}$  & $0.1$  & \cite{wolf2007kinetic} \\
 $K_{A,IC}$    & $IC$ half saturation coeff. for $A$ &  $kmol(IC) \ m^{-3}$ & $10^{-4}$ & \cite{wolf2007kinetic} \\
  $K_{A,DOC}$ & $DOC$ half saturation coeff. for $A$ & $kg(COD) \ m^{-3}$ & $5 \cdot10^{-3}$ & \cite{wolf2007kinetic} \\
 $K_{A,NO3}$ & $NO_3$ half saturation coeff. for $A$  &  $kmol(NO_3) \ m^{-3}$ & $1.2\cdot10^{-6}$ & \cite{wolf2007kinetic} \\
 $K_{A,NH3}$ & $NH_3$ half saturation coeff. for $A$ &  $kmol(NH_3) \ m^{-3}$ & $1.2\cdot10^{-6}$ &\cite{wolf2007kinetic} \\
 $K_{A,I}$ & Light inhibition coefficient for $A$ & $kmol(e^-) \ m^{-2} \ d^{-1}$ & $8\cdot10^{-5}$ & \cite{wolf2007kinetic} \\
 $K_{C,IC}$    & $IC$ half saturation coeff. for $C$ &  $kmol(IC) \ m^{-3}$ & $10^{-4}$ & \cite{wolf2007kinetic} \\
 $K_{C,DOC}$ & $DOC$ half saturation coeff. for $C$ & $kg(COD) \ m^{-3}$ & $5 \cdot10^{-3}$ & \cite{wolf2007kinetic} \\
 $K_{C,NO3}$ & $NO_3$ half saturation coeff. for $C$  &  $kmol(NO_3) \ m^{-3}$ & $1.2\cdot10^{-6}$ & \cite{wolf2007kinetic} \\
 $K_{C,NH3}$ & $NH_3$ half saturation coeff. for $C$ &  $kmol(NH_3) \ m^{-3}$ & $1.2\cdot10^{-6}$ &\cite{wolf2007kinetic} \\
 $K_{C,I}$ & Light inhibition coefficient for $C$ & $kmol(e^-) \ m^{-2} \ d^{-1}$ & $8\cdot10^{-5}$ & \cite{wolf2007kinetic} \\
 $K_{H,DOC}$ & $DOC$ half saturation coeff. for $H$ & $kg(COD) \ m^{-3}$ & $4 \cdot10^{-3}$ & \cite{wolf2007kinetic} \\
 $K_{H,NO3}$ & $NO_3$ half saturation coeff. for $H$  &  $kmol(NO_3) \ m^{-3}$ & $3.6\cdot10^{-5}$ & \cite{wolf2007kinetic} \\
 $K_{H,NH3}$ & $NH_3$ half saturation coeff. for $H$ &  $kmol(NH_3) \ m^{-3}$ & $3.6\cdot10^{-6}$ & \cite{henze2000activated} \\
  $K_{H,O2}$ & $O_2$ half saturation coeff. for $H$ & $kmol(O_2) \ m^{-3}$ & $6.25\cdot10^{-6}$ & \cite{wolf2007kinetic} \\
 $K_{N,IC}$    & $IC$ half saturation coeff. for $N$ &  $kmol(IC) \ m^{-3}$ & $10^{-4}$ & \cite{wolf2007kinetic} \\
 $K_{N,NH3}$ & $NH_3$ half saturation coeff. for $N$ &  $kmol(NH_3) \ m^{-3}$ & $7\cdot10^{-5}$ &\cite{wolf2007kinetic} \\
  $K_{N,O2}$ & $O_2$ half saturation coeff. for $N$ & $kmol(O_2) \ m^{-3}$ & $1.56\cdot10^{-5}$ & \cite{wolf2007kinetic} \\
  $K^{in}_{O2,max}$ & Max inhibition coefficient of $O_2$ on $A$ and $C$     & 
 $kmol(O_2) \ m^{-3}$ & $10^{-3}$     &   \cite{li2016investigating}    \\ 
  $K_{R_{CO2/O2}}$ & Half saturation coeff. for $O_2$ inhibition     &  $--$ & $0.35$     &   \cite{li2016investigating}    \\ 
 $Y_{H}$ & Yield of $H$ on $DOC$  & $kg(COD) \ kg(COD)^{-1}$ & $0.63$ & \cite{wolf2007kinetic} \\ 
 $Y_{N}$ & Yield of $N$ on $NO_3$  & $kg(COD) \ kg(NO_3-N)^{-1}$ & $0.24$ & \cite{wolf2007kinetic}  \\
 $Y_{DOC}$ & Yield of $A$ and $C$ on $DOC$  & $kg(COD) \ kg(COD)^{-1}$ & $0.5$ & (a) \\  
 \hline
 \multicolumn{5}{l}{(a) Assumed}\\
 
 \end{tabular}
 \caption{Kinetic parameters} \label{t5.3.1}
 \end{center}
 \end{spacing}
 \end{scriptsize}
 \end{table}

  \begin{table}[ht]
\begin{scriptsize}
\begin{spacing}{1.2}
 \begin{center}
 \begin{tabular}{llccc}
 \hline
{\textbf{Parameter}} & {\textbf{Definition}} & {\textbf{Unit}} & {\textbf{Value}} & {\textbf{Ref}}
 \\
 \hline
  $\phi_{EPS,A}$ & Relative rate of EPS formation to $A$ production  & $--$ & $0.1$ & (a)  \\
 $\phi_{EPS,C}$ & Relative rate of EPS formation to $C$ production  & $--$ & $0.3$ & (a)  \\
 $k_{EPS,H}$ & EPS fraction produced by $H$  & $--$ & $0.18$ & \cite{merkey2009modeling}  \\
 $k_{EPS,N}$ & EPS fraction produced by $N$  & $--$ & $0.075$ & \cite{merkey2009modeling}  \\
 $k_{DOC}$ & DOC release fraction by $A$ and $C$  & $--$ & $0.05$ & \cite{tenore94modelling}\\
 $k_{La}$ & $O_2$ mass transfer coefficient & $d^{-1}$ & $23.3$ & \cite{munoz2014modeling} \\
  $S_{O2,sat}$ & $O_2$ saturation concentration in bulk liquid & $kmol(O_2) \ m^{-3}$ & $2.4\cdot10^{-4}$ & \cite{munoz2014modeling} \\
 $I_{opt,A}$    &   Optimum light intensity for $A$                   & $kmol (e^-) \ m^{-2} \ d^{-1}$ & $0.01728$   &   \cite{flora1995modeling}   \\
 $I_{opt,C}$    &   Optimum light intensity for $C$                   & $kmol (e^-) \ m^{-2} \ d^{-1}$ & $0.00864$   &   (a)   \\ 
 $\eta_{A}$     & Coeff. of adaptability to non-optimal light for $A$  & $--$ & $1$  &    (a)  \\
 $\eta_{C}$      & Coeff. of adaptability to non-optimal light for $C$   & $--$ & $0.6$  &    (a)  \\
 $k_{tot}$       & Light attenuation coefficient             & $m^{2} \ kg^{-1}$ & $210$   &   \cite{wolf2007kinetic}  \\
 $D_{S,IC}$ & Diffusion coefficient of $IC$ in biofilm & 
 $m^2 \ d^{-1}$ & $1.32\cdot10^{-4}$ & \cite{wolf2007kinetic} \\
 $D_{S,DOC}$ & Diffusion coefficient of $DOC$ in biofilm & $m^2 \ d^{-1}$ & $0.83\cdot10^{-4}$ & \cite{wanner1986multispecies} \\
 $D_{S,NO3}$ & Diffusion coefficient of $NO_3$ in biofilm & $m^2 \ d^{-1}$ & $1.18\cdot10^{-4}$ &  \cite{wolf2007kinetic} \\
 $D_{S,NH3}$ & Diffusion coefficient of $NH_3$ in biofilm & $m^2 \ d^{-1}$ & $1.49\cdot10^{-4}$ & \cite{wanner1986multispecies} \\
 $D_{S,O2}$ & Diffusion coefficient of $O_2$ in biofilm & $m^2 \ d^{-1}$ & $1.75\cdot10^{-4}$ & \cite{wanner1986multispecies} \\
  $v_{a,C}$    & Attachment velocity of $\psi^*_{C}$  & $m \ d^{-1}$ & $5\cdot10^{-3}$    &    (a) \\
   $v^0_{a,A}$    & Attachment velocity of $\psi^*_{A}$ & $m \ d^{-1}$ & $5\cdot10^{-4}$    &    (a)    \\
   $v^0_{a,H}$    & Attachment velocity of $\psi^*_{H}$ & $m \ d^{-1}$ & $5\cdot10^{-4}$    &    (a)    \\
   $v^0_{a,N}$   & Attachment velocity of $\psi^*_{N}$ & $m \ d^{-1}$  & $5\cdot10^{-4}$    &    (a)    \\
      $K_{C}$   & Half saturation coeff. of $\psi^*_{C}$ on $\psi^*_{A}$, $\psi^*_{H}$,$\psi^*_{N}$ attachment & $kg (COD) \ m^{-3}$  & $3\cdot10^{-2}$    &    (a)    \\
 $\rho$    & Biofilm density  & $kg(COD) \ m^{-3}$ & $37$  & \cite{munoz2014modeling}  \\
 $\lambda$    & Constant detachment coefficient   &  $m^{-1} \ d^{-1}$ & 50  &    (a)   \\
  $V$  & Reactor volume  &  $m^{3}$   & 400  &    (a)   \\
  $N_G$  &  Number of granules in the reactor &  $--$   & $2.4\cdot10^{10}$ & (a)   \\
  $\tau$  &  Duration of the cycle &  $d$   & $0.25$ & (a)   \\
  $\gamma$  &  Fraction of suspended biomass lost during the emptying &  $--$   & $0.2$ & (a)   \\
  $\omega$  &  Emptying/refilling ratio &  $--$   & $0.5$ & (a)   \\
 \hline
 \multicolumn{5}{l}{(a) Assumed}\\
 
  \end{tabular}
 \caption{Other model parameters} \label{t5.3.2}
 \end{center}
 \end{spacing}
 \end{scriptsize}
 \end{table}

 The treatment process takes place in an SBR reactor, where the wastewater is fed discontinuously and treated through a cycle which is repeated over time. The duration of each cycle has been set to six hours ($\tau=0.25 \ d$), as in the experiments carried out in \cite{abouhend2018oxygenic}. Each cycle is based on a first phase of darkness (three hours), when no light source is provided to the system, followed by a phase of light (three hours) when the reactor is supposed to be homogeneously illuminated. In particular, in the first two studies, the incident light intensity $I_0$ is fixed at $0.008 \ kmol \ m^{-2} \ d^{-1}$, similar to \cite{abouhend2018oxygenic,downes2019success}. As noted in \cite{abouhend2018oxygenic}, this value favors the growth of cyanobacteria at the expense of microalgae, and therefore guarantees greater chances of success of the granulation process. Instead, different values of $I_0$ have been used in the last study and they will be specified later. Cyanobacteria play a key role in the formation of photogranules thanks to their properties. Such role is modelled by setting the attachment velocity of cyanobacteria an order of magnitude higher than the attachment velocity of the other species. The number of granules $N_{G}$ is calculated in order to have a reactor filled by biofilm biomass for approximately $25\%$ of volume once the granules have reached the steady-state dimension. All parameters used in this model are reported in Tables \ref{t5.3.1} and \ref{t5.3.2}.
 
The same composition is considered for the influent of each treatment cycle. In particular, no suspended biomass is supposed to be present in the influent, while the concentration of soluble substrates varies from case to case and will be specified in each study as appropriate. The initial concentration of soluble substrates within the reactor has been set equal to the concentration within the influent. Furthermore, a phototrophic inoculum of suspended cyanobacteria and microalgae is considered, where suspended heterotrophic and nitrifying bacteria are present in smaller amounts: $\psi^*_{A,0} = \psi^*_{C,0} = 300 \ g \  m^{-3}$, $\psi^*_{H,0} = \psi^*_{N,0} = 50 \ g \  m^{-3}$.

\subsection{Study 1 - OPGs-based system fed with a municipal wastewater} \label{n5.3.1}
\begin{figure*}
 \fbox{\includegraphics[width=1\textwidth, keepaspectratio]{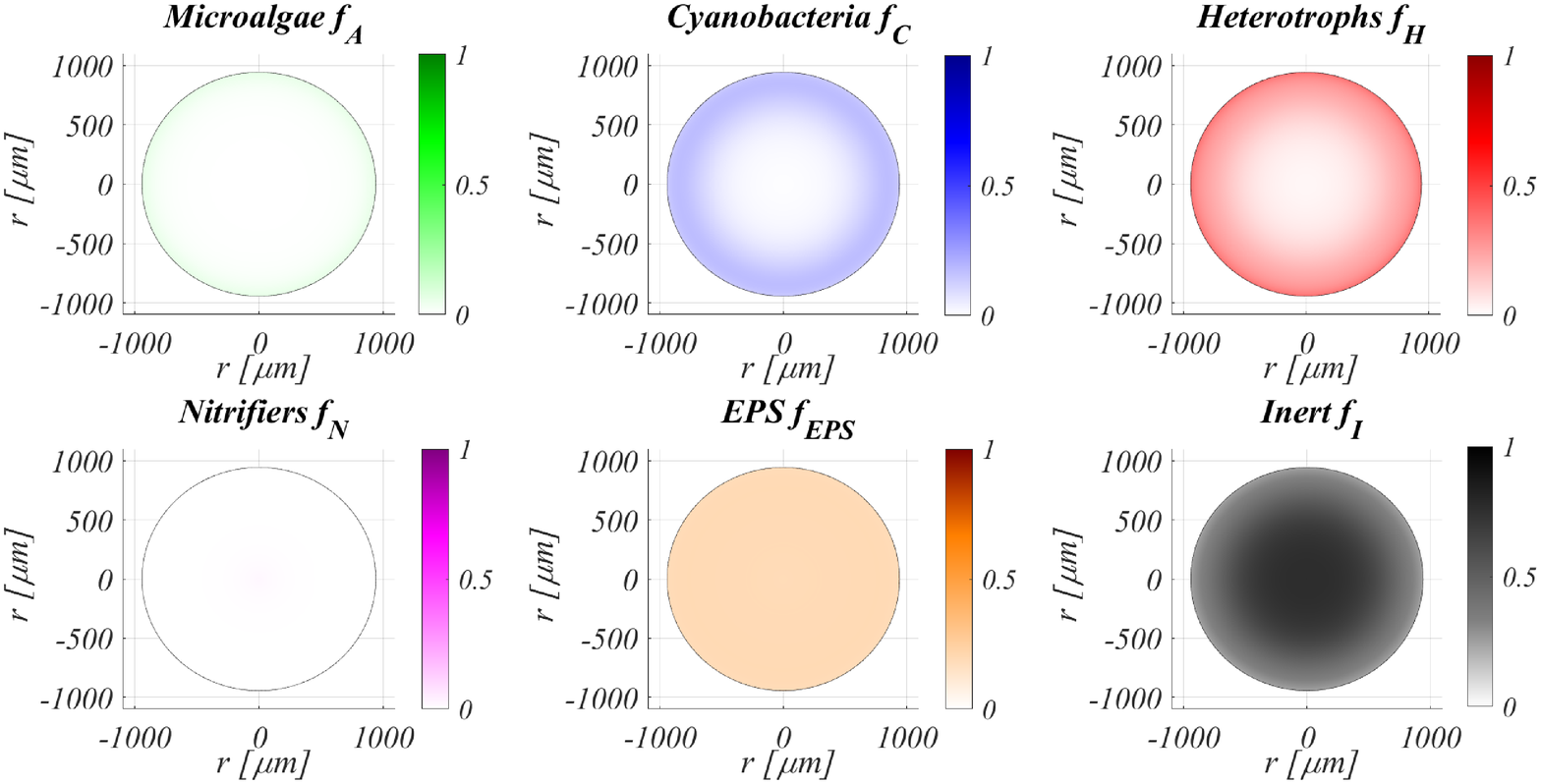}}   
 \caption{Study 1 - Microbial species distribution in the diametrical section, at $T = 50 \ d$. Wastewater influent composition: $S^{in}_{IC} = 180 \ g \  m^{-3}$ (inorganic carbon), $S^{in}_{DOC} = 500 \ g \  m^{-3}$ (organic carbon), $S^{in}_{NH_3} = 50 \ g \  m^{-3}$ (ammonia), $S^{in}_{NO_3} = 0 \ g \  m^{-3}$ (nitrate), $S^{in}_{O_2} = 0 \ g \  m^{-3}$ (oxygen). Incident light intensity: $I_0 = 0.008 \ kmol \ m^{-2} \ d^{-1}$.}
          \label{f5.4.1.1} 
 \end{figure*}

\begin{figure*}
 \fbox{\includegraphics[width=1\textwidth, keepaspectratio]{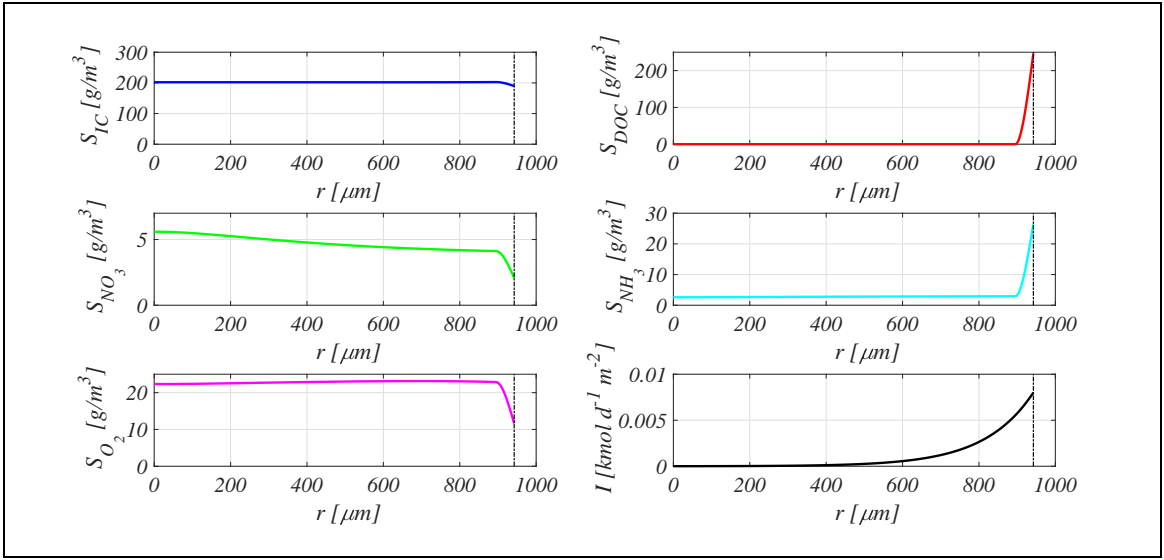}}   
 \caption{Study 1 - Distribution of soluble substrates and trend of light intensity along the granule radius at $T=50 \ d$ (before the emptying and refilling). Wastewater influent composition: $S^{in}_{IC} = 180 \ g \  m^{-3}$ (inorganic carbon), $S^{in}_{DOC} = 500 \ g \  m^{-3}$ (organic carbon), $S^{in}_{NH_3} = 50 \ g \  m^{-3}$ (ammonia), $S^{in}_{NO_3} = 0 \ g \  m^{-3}$ (nitrate), $S^{in}_{O_2} = 0 \ g \  m^{-3}$ (oxygen). Incident light intensity: $I_0 = 0.008 \ kmol \ m^{-2} \ d^{-1}$.}
          \label{f5.4.1.2} 
 \end{figure*}

\begin{figure*}
 \fbox{\includegraphics[width=1\textwidth, keepaspectratio]{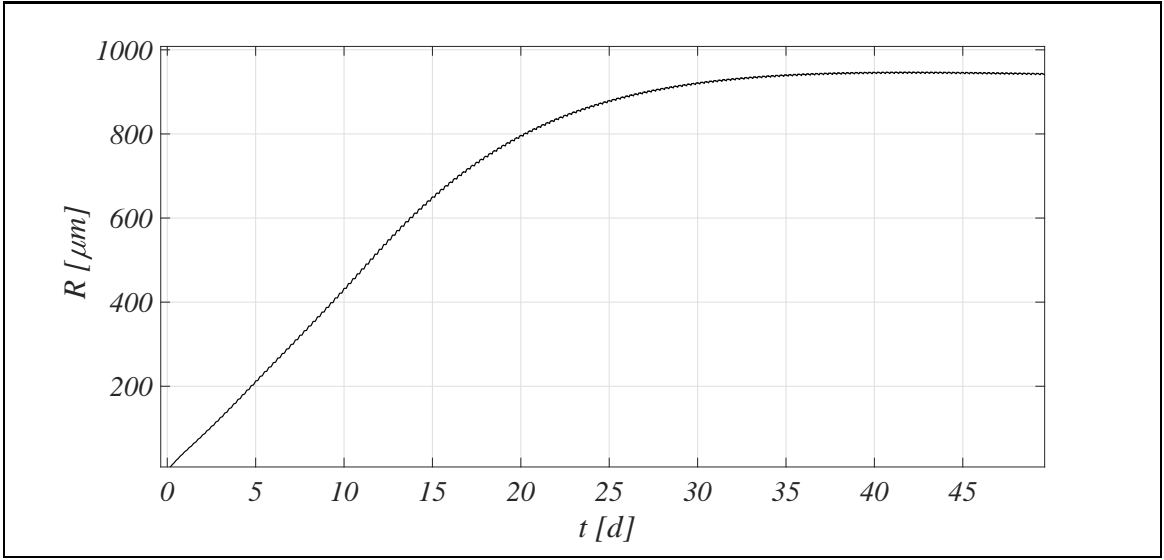}}   
 \caption{Study 1 - Evolution of biofilm radius over time. Wastewater influent composition: $S^{in}_{IC} = 180 \ g \  m^{-3}$ (inorganic carbon), $S^{in}_{DOC} = 500 \ g \  m^{-3}$ (organic carbon), $S^{in}_{NH_3} = 50 \ g \  m^{-3}$ (ammonia), $S^{in}_{NO_3} = 0 \ g \  m^{-3}$ (nitrate), $S^{in}_{O_2} = 0 \ g \  m^{-3}$ (oxygen). Incident light intensity: $I_0 = 0.008 \ kmol \ m^{-2} \ d^{-1}$.}
          \label{f5.4.1.3} 
 \end{figure*}

\begin{figure*}
 \fbox{\includegraphics[width=1\textwidth, keepaspectratio]{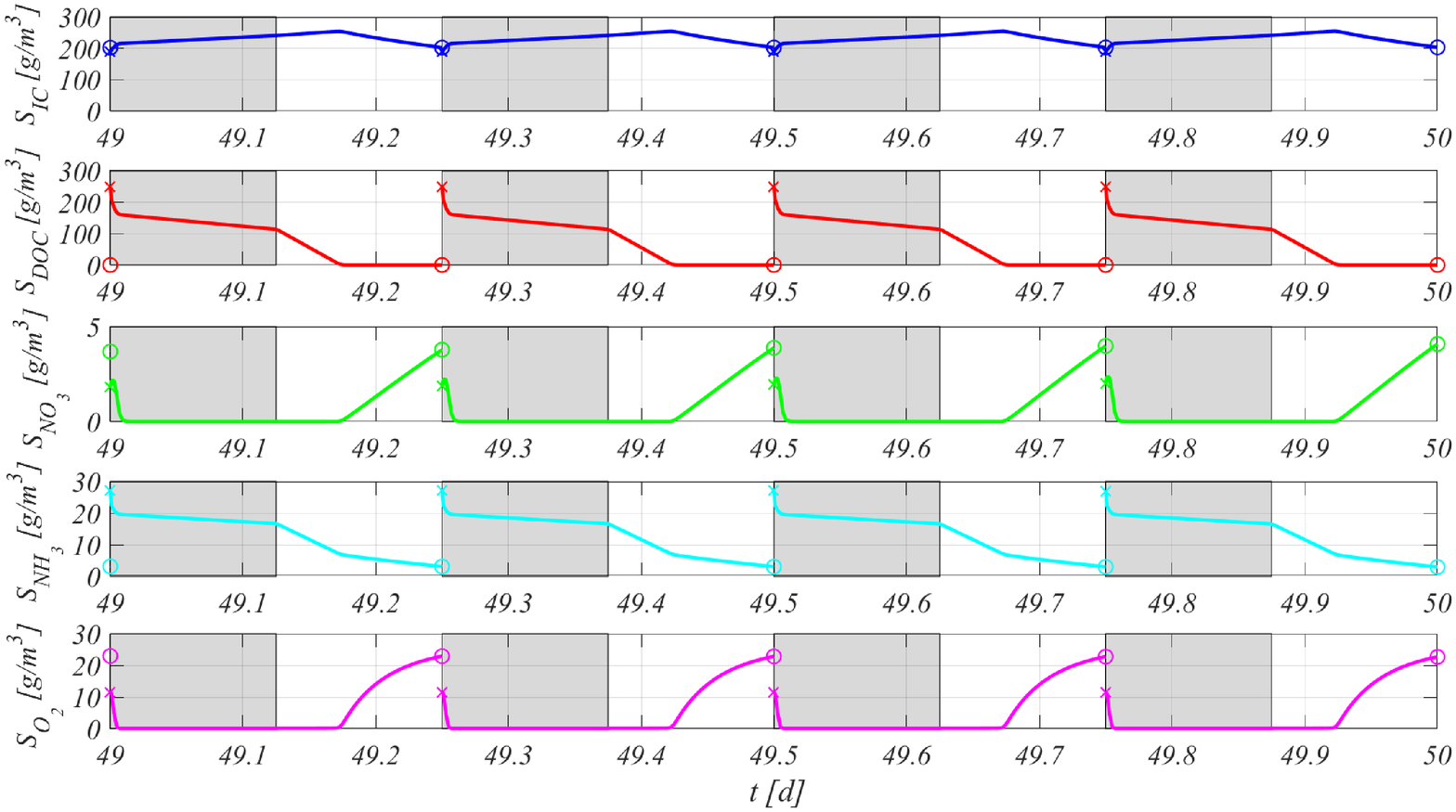}}   
 \caption{Study 1 - Evolution of soluble substrates concentration within the reactor, from $T=49 \ d$ to $T=50 \ d$ (four consecutive six-hours treatment cycles). Wastewater influent composition: $S^{in}_{IC} = 180 \ g \  m^{-3}$ (inorganic carbon), $S^{in}_{DOC} = 500 \ g \  m^{-3}$ (organic carbon), $S^{in}_{NH_3} = 50 \ g \  m^{-3}$ (ammonia), $S^{in}_{NO_3} = 0 \ g \  m^{-3}$ (nitrate), $S^{in}_{O_2} = 0 \ g \  m^{-3}$ (oxygen). Incident light intensity: $I_0 = 0.008 \ kmol \ m^{-2} \ d^{-1}$. Grey portions indicate the dark phases, white portions indicate the light phases.}
          \label{f5.4.1.4} 
 \end{figure*}

\begin{figure*}
 \fbox{\includegraphics[width=1\textwidth, keepaspectratio]{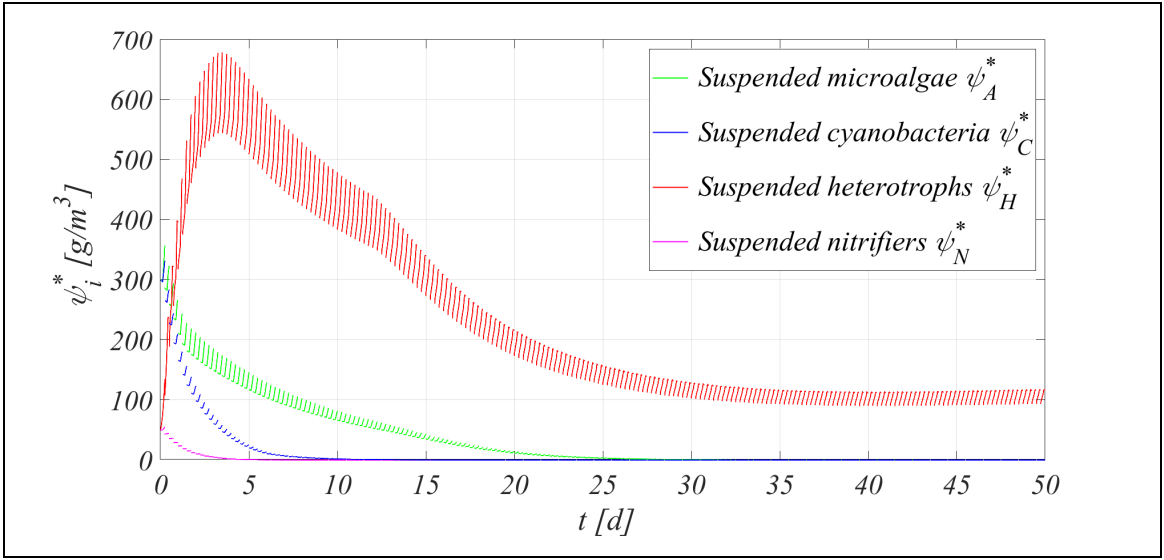}}   
 \caption{Study 1 - Evolution of suspended biomasses concentration within the reactor. Wastewater influent composition: $S^{in}_{IC} = 180 \ g \  m^{-3}$ (inorganic carbon), $S^{in}_{DOC} = 500 \ g \  m^{-3}$ (organic carbon), $S^{in}_{NH_3} = 50 \ g \  m^{-3}$ (ammonia), $S^{in}_{NO_3} = 0 \ g \  m^{-3}$ (nitrate), $S^{in}_{O_2} = 0 \ g \  m^{-3}$ (oxygen). Incident light intensity: $I_0 = 0.008 \ kmol \ m^{-2} \ d^{-1}$.}
          \label{f5.4.1.5} 
 \end{figure*}

\
In this study, the microbial composition of the photogranules and the treatment process within the OPG-based system are investigated in the case of a typical municipal wastewater. For this purpose, the following influent composition is considered: $S^{in}_{IC} = 180 \ g \  m^{-3}$, $S^{in}_{DOC} = 500 \ g \  m^{-3}$, $S^{in}_{NH_3} = 50 \ g \  m^{-3}$, $S^{in}_{NO_3} = S^{in}_{O_2} = 0 \ g \  m^{-3}$. The results are shown in Figs. \ref{f5.4.1.1}-\ref{f5.4.1.5}.

Fig. \ref{f5.4.1.1} reports the distribution of biomass within the photogranules, after 50 days of simulation, while Fig. \ref{f5.4.1.2} presents the distribution of soluble substrates and the trend of light intensity along the granule radius at $T=50 \ d$ (before the emptying and refilling). From Fig. \ref{f5.4.1.1}, it can be seen that photogranules are composed of a significant fraction of cyanobacteria, expecially in the external layers due to optimal light conditions, while their concentration reduces in the interior part, where light intensity decreases due to attenuation phenomena (see Fig. \ref{f5.4.1.2}). On the other hand, microalgae are present in small amounts and confined to the outermost layers. In addition, optimal conditions for the growth of heterotrophic bacteria occur: the influent is rich in $DOC$ and phototrophic organisms (microalgae and cyanobacteria) produce large amounts of $O_2$. As well as cyanobacteria, the fraction of heterotrophic bacteria is high in the external layers and reduces toward the granule nucleus, due to the $DOC$ concentration gradients occurring along the granules (Fig. \ref{f5.4.1.2}). Indeed, the $DOC$ concentration reduces along the granule due to diffusion and consumption in the outermost layers. Nitrifying bacteria are almost absent, because have lower maximum growth rates than heterotrophic bacteria which, consequently, in presence of $DOC$ are more competitive in the use of $O_2$. All active species contribute to the secretion of $EPS$, which has a homogeneous distribution throughout the granule. Finally, large quantities of inert material deriving by decay processes are found in the granule, especially in the internal layers.

Fig. \ref{f5.4.1.3} reports the evolution of the photogranule radius $R(t)$ over time. In the initial phase the intense attachment process lead to a linear increase of the radius over time. Subsequently, detachment processes become more relevant as the granule dimension increases, and limit the growth of the granule, which reaches its steady-state dimension after 40-50 days.

Fig. \ref{f5.4.1.4} shows how the concentration of substrates within the reactor varies over time. The observation period reported goes from day $49$ to day $50$, which is likely to describe the treatment process under operating conditions. Indeed, it is observed that the start-up of the system is completed before this time: photogranules have reached a steady-state dimension and the trend of the substrates concentration and the final composition of the effluent are repeated identically in each cycle. Since all cycles last six hours, Fig. \ref{f5.4.1.4} shows four consecutive cycles, where a dark phase (grey portion of the graphic) and a light phase (white portion of the graphic) are distinguished. At the end of each cycle there is a discontinuity in the graphs, which indicates the emptying of the reactor and refilling with a new wastewater volume to be treated. At the beginning of the cycle, a very rapid reduction of $DOC$ and $NH_3$ is observed, due to all aerobic processes of microbial growth. In particular, the large amount of $O_2$ produced in the previous cycle leads to optimal conditions for the growth of aerobic heterotrophic bacteria, nitrifying bacteria, microalgae and cyanobacteria (in absence of light the last two species adopt heterotrophic metabolic pathways). When $O_2$ concentration reduces, anoxic heterotrophs grow by consuming the small amount of $NO_3$ remained within the reactor from the previous cycle. However, the amount of $NO_3$ is extremely low and therefore the contribution that anoxic processes provide to the consumption of $DOC$ and $NH_3$ is pratically negligible. When oxidized substrates drop to very low concentrations, they become limiting for microbial kinetics and a clear change in slope is observed in the trends of $DOC$, $NH_3$ and $IC$: growth rates are very low and lead to a slight consumption of $DOC$ and $NH_3$ and a slight production of $IC$. This trend marks the remaining dark period. After it, the light source is provided in the reactor. Both cyanobacteria and microalgae adopt photoautotrophic metabolic strategies and produce large amounts of $O_2$. Aerobic heterotrophic and nitrifying bacteria find again optimal conditions for their metabolic growth and consequently, the consumption rates of $DOC$ and $NH_3$ increases. As long as $DOC$ is present in the reactor, heterotrophic bacteria govern the treatment process, being more competitive than nitrifying bacteria in using the $O_2$ produced by cyanobacteria and microalgae. To confirm this, no production of $NO_3$ is observed. When $DOC$ concentration drops, the growth of the heterotrophs decreases, their consumption of $O_2$ and $NH_3$ reduces, and nitrifying bacteria take over the process. Indeed, $NH_3$ continues to be consumed with a different rate and a slight production of $NO_3$ is observed. Anyway, as already observed in Fig. \ref{f5.4.1.1}, the fraction of nitrifying bacteria within the granule and their growth rate are very low, therefore the production of $NO_3$ is limited and the $O_2$ concentration increases again because the autotrophic consumption does not balance the photoautotrophic production. Regarding the $IC$ concentration, it is associated to the trend of $DOC$: when $DOC$ is present in the reactor, $IC$ increases over time, because heterotrophic bacteria are the most competitive species and their $IC$ production prevails over consumption processes, while $IC$ reduces over time when $DOC$ finishes, because the heterotrophic growth is inhibited and the photosynthetic activity of cyanobacteria and microalgae prevails. As can be seen in the Fig. \ref{f5.4.1.4}, the trends of the substrates concentration described above are repeated identically in each cycle and are representative of the treatment cycle of an OPG-based system. The substrates concentrations at the end of the cycle are representative of the effluent composition. In this case, the removal efficiency is high: $DOC$ has been completely removed and very low concentrations of nitrogen compounds are observed (less than $5 \ g \  m^{-3}$ of $NO_3$). Obviously the operation of this system and the evolution of the granules are strongly influenced by the type of wastewater influent. For this reason, in the second study the treatment process will be investigated in the case of different compositions of the influent.

Fig. \ref{f5.4.1.5} reports the concentration of suspended biomasses within the reactor over time. It should be noted that the concentration trends have a discontinuity every six hours, at the end of each cycle, when $20\%$ of each suspended biomass is supposed to be lost during the emptying phase, due to not perfect settling properties. Consequently, phenomena leading to the reduction of suspended biomass within the SBR are the attachment (the swich of phenotype from suspended to sessile), the emptying of the reactor and the decay processes, while the processes of metabolic growth lead to an increase of the suspended biomass. 

 Heterotrophic bacteria have higher maximum growth rates than other microbial species and their concentration significantly increases during the start-up phase of the reactor, when the granules are small and higher amounts of substrates are available for the suspended biomass. Later, photogranules grow and the contribution of sessile biomass to the consumption of soluble substrates increases. As a result, the amount of substrates available for suspended biomass reduces and the concentration of suspended heterotrophs decreases cycle after cycle until to reach a final trend which is repeated identically over time. Other species have low maximum growth rates and are all washed out over time, due to the processes described above. Suspended nitrifying bacteria are the slowest species to grow due to their low maximum growth rate and to the competition with heterotrophs for $O_2$, and are the first to be washed out. Suspended cyanobacteria and microalgae have similar metabolic activities but the first have better attachment properties and the higher attachment flux explains the faster reduction observed.

\subsection{Study 2 - Effects of wastewater influent composition on the process} \label{n5.3.2}
\begin{figure*}
 \fbox{\includegraphics[width=1\textwidth, keepaspectratio]{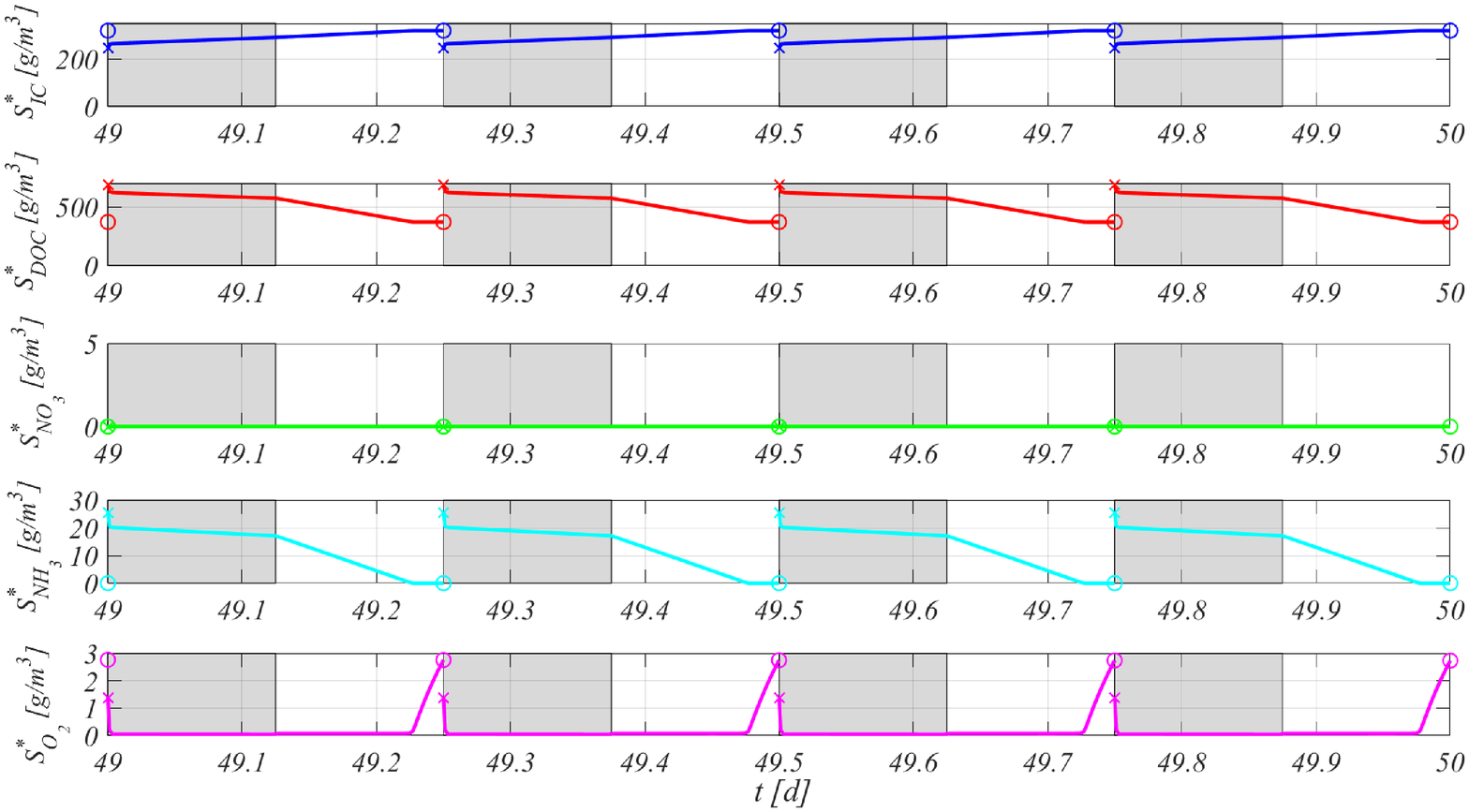}}   
 \caption{Study 2 (Case 1) - Evolution of soluble substrates concentration within the reactor, from $T=49 \ d$ to $T=50 \ d$ (four consecutive six-hours treatment cycles). Wastewater influent composition: $S^{in}_{IC} = 180 \ g \  m^{-3}$ (inorganic carbon), $S^{in}_{DOC} = 1000 \ g \  m^{-3}$ (organic carbon), $S^{in}_{NH_3} = 50 \ g \  m^{-3}$ (ammonia), $S^{in}_{NO_3} = 0 \ g \  m^{-3}$ (nitrate), $S^{in}_{O_2} = 0 \ g \  m^{-3}$ (oxygen). Incident light intensity: $I_0 = 0.008 \ kmol \ m^{-2} \ d^{-1}$. Grey portions indicate the dark phases, white portions indicate the light phases.}
          \label{f5.4.2.1} 
 \end{figure*}

\begin{figure*}
 \fbox{\includegraphics[width=1\textwidth, keepaspectratio]{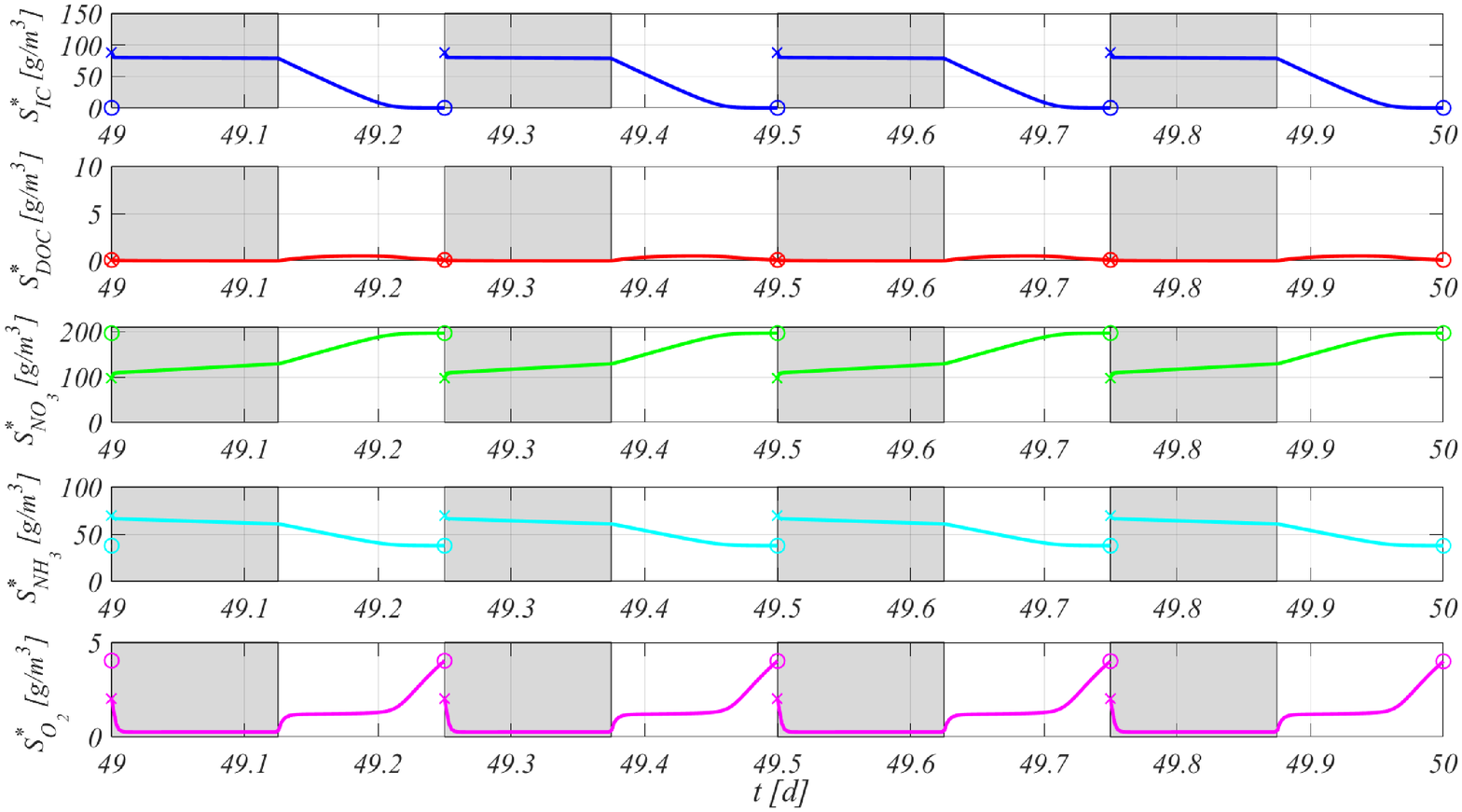}}   
 \caption{Study 2 (Case 2) - Evolution of soluble substrates concentration within the reactor, from $T=49 \ d$ to $T=50 \ d$ (four consecutive six-hours treatment cycles). Wastewater influent composition: $S^{in}_{IC} = 180 \ g \  m^{-3}$ (inorganic carbon), $S^{in}_{DOC} = 0 \ g \  m^{-3}$ (organic carbon), $S^{in}_{NH_3} = 100 \ g \  m^{-3}$ (ammonia), $S^{in}_{NO_3} = 0 \ g \  m^{-3}$ (nitrate), $S^{in}_{O_2} = 0 \ g \  m^{-3}$ (oxygen). Incident light intensity: $I_0 = 0.008 \ kmol \ m^{-2} \ d^{-1}$. Grey portions indicate the dark phases, white portions indicate the light phases.}
          \label{f5.4.2.2} 
 \end{figure*}

\begin{figure*}
 \fbox{\includegraphics[width=1\textwidth, keepaspectratio]{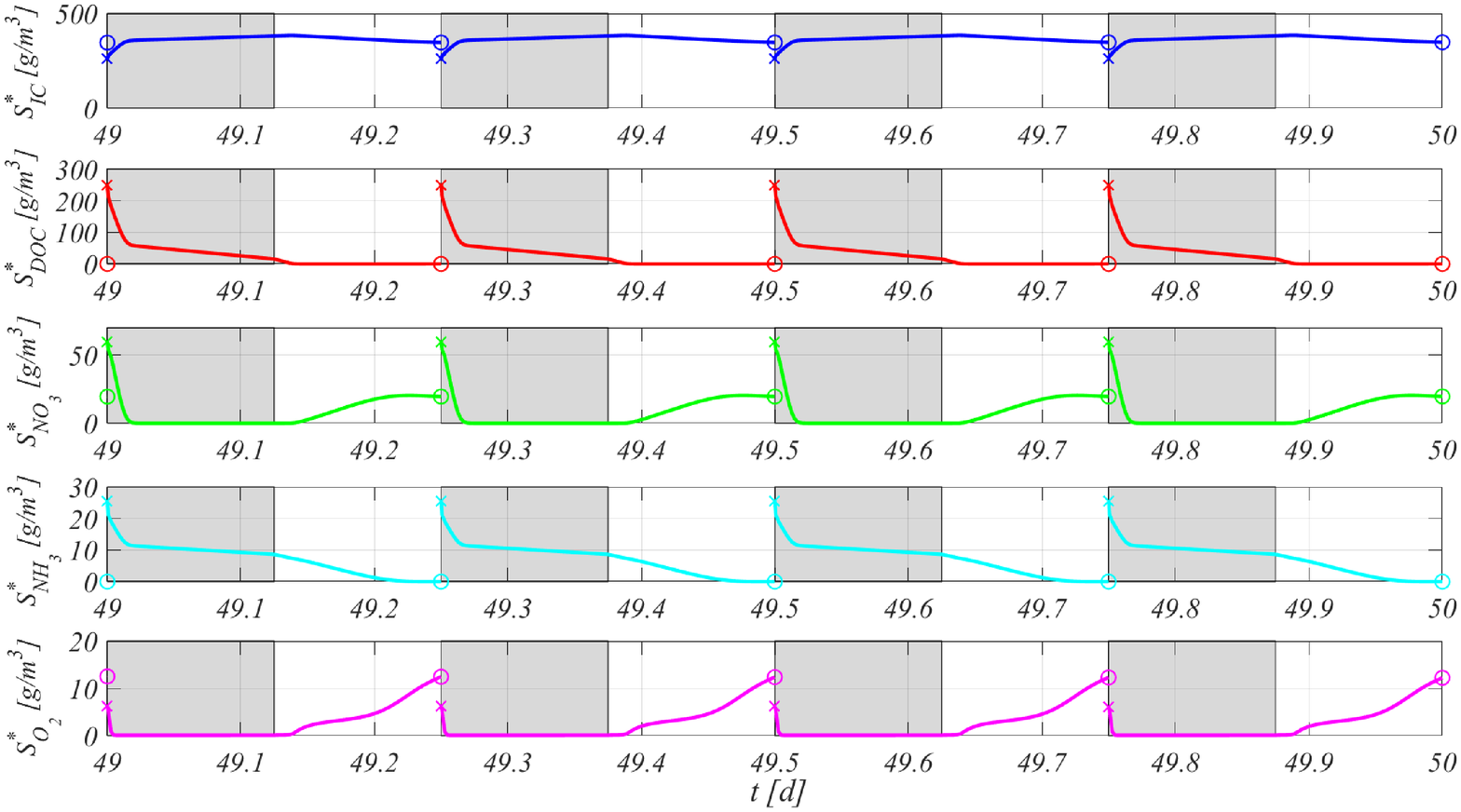}}   
 \caption{Study 2 (Case 3) - Evolution of soluble substrates concentration within the reactor, from $T=49 \ d$ to $T=50 \ d$ (four consecutive six-hours treatment cycles). Wastewater influent composition: $S^{in}_{IC} = 180 \ g \  m^{-3}$ (inorganic carbon), $S^{in}_{DOC} = 500 \ g \  m^{-3}$ (organic carbon), $S^{in}_{NH_3} = 50 \ g \  m^{-3}$ (ammonia), $S^{in}_{NO_3} = 100 \ g \  m^{-3}$ (nitrate), $S^{in}_{O_2} = 0 \ g \  m^{-3}$ (oxygen). Incident light intensity: $I_0 = 0.008 \ kmol \ m^{-2} \ d^{-1}$. Grey portions indicate the dark phases, white portions indicate the light phases.}
          \label{f5.4.2.3} 
 \end{figure*}

\begin{figure*}
 \fbox{\includegraphics[width=1\textwidth, keepaspectratio]{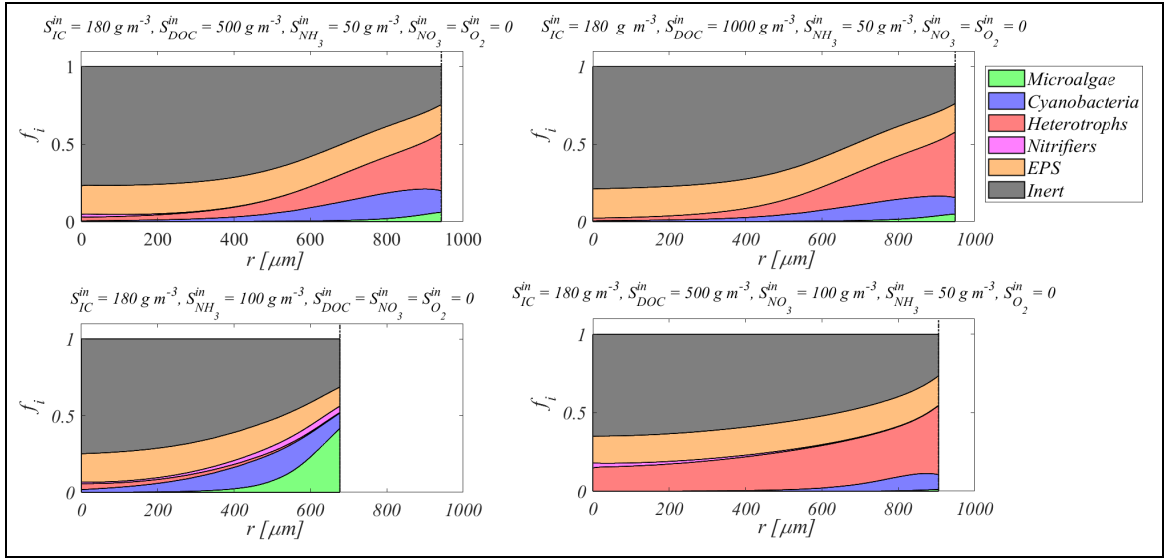}}   
 \caption{Study 2 - Microbial species distribution across the radius of the granule at $T = 50 \ d$, in the four cases reported in Study 1 and Study 2. Incident light intensity: $I_0 = \ 0.008 \ kmol \ m^{-2} \ d^{-1}$.}
          \label{f5.4.2.4} 
 \end{figure*}

\
The performances of the OPG-based system are strongly influenced by the composition of the wastewater influent, which governs the evolution and the microbial composition of photogranules and consequently, the treatment process. The carbon and nitrogen loads and their ratio affect the metabolic activity of microbial species populating the granules and the purifying ability of these biomass assemblies. Therefore, it is interesting to compare the results of the previous study with other case study, where different types of influent wastewater are considered: an higher strength carbon wastewater (case 1), an ammonia wastewater (case 2) and a municipal wastewater with the presence of nitrate (case 3). The results of this study are summarized in Figs. \ref{f5.4.2.1}-\ref{f5.4.2.4}. In particular, Figs. \ref{f5.4.2.1}-\ref{f5.4.2.3} focus on the concentration of soluble substrates in the system in the three cases, while the microbial composition of photogranules in such cases is reported in Fig. \ref{f5.4.2.4}, and compared with the results of the first study.

Fig. \ref{f5.4.2.1} reports the concentration of soluble substrates within the reactor over time, in the period between 49 and 50 days, in the case of a high strength organic wastewater (case 1: $S^{in}_{IC} = 180 \ g \  m^{-3}$, $S^{in}_{DOC} = 1000 \ g \  m^{-3}$, $S^{in}_{NH_3} = 50 \ g \  m^{-3}$, $S^{in}_{NO_3} = S^{in}_{O_2} = 0$). As in the case presented in the previous study, at the beginning of the cycle there is a rapid heterotrophic consumption of $O_2$ which is still in the reactor from the previous cycle. When this concentration approaches zero, the trend of substrate concentrations does not show high variations until the end of the dark period: a slight reduction of $DOC$ and $NH_3$ concentration and a slight increase of $IC$ concentration are observed, caused by the metabolic activities of aerobic heterotrophic bacteria, microalgae and cyanobacteria. When the light period begins, cyanobacteria and microalgae carry out their photosynthetic activity and produce $O_2$ necessary for heterotrophs. As a result, a clear change in slope is observed in the trends of $DOC$ and $NH_3$ concentrations, which reduce due to the fast consumption by heterotrophic bacteria growing in optimal conditions. In the last part of the cycle, $NH_3$ runs out, hence, the heterotrophic kinetics slow down again and the $O_2$ concentration increases up to $3  \ g \  m^{-3}$. As can be seen, in this case $NO_3$ concentration is nearly zero throughout the cycle and suggests that the amount of nitrifying bacteria populating the photogranule is negligible. Indeed, as already mentioned in Study 1, compared to heterotrophs, nitrifying bacteria have lower maximum growth rates and are less competitive in the use of $O_2$, and develop only in poor-DOC environments. In this case the high $DOC$ amounts present throughout the cycle do not allow their growth. At the end of the cycle, although $NH_3$ has been completely removed, large concentrations of $DOC$ are still present in the effluent.

Fig. \ref{f5.4.2.2} presents the trend of the substrates concentration within the reactor in the case of an ammonia wastewater influent (case 2: $S^{in}_{IC} = 180 \ g \  m^{-3}$, $S^{in}_{NH_3} = 100 \ g \  m^{-3}$, $S^{in}_{DOC} = S^{in}_{NO_3} = S^{in}_{O_2} = 0$). In this case, the process is strongly influenced by the absence of $DOC$ in the influent. Indeed, $DOC$ concentration is extremely low and limiting for the heterotrophic growth throughout the cycle, and the process is mainly driven by nitrifying bacteria, microalgae and cyanobacteria. In dark conditions, the growth rate of microalgae, cyanobacteria and nitrifying bacteria is very low due to the $DOC$ and $O_2$ shortage. For this reason, the concentration of $DOC$, $IC$ and $O2$ is almost constant throughout the dark period and only a slight consumption of $NH_3$ and a slight production of $NO_3$ are observed. As light conditions within the system change, microbial growth rates increase. In particular, photoautotrophic processes in presence of light lead to the production of $O_2$; consequently the growth rate of nitrifying bacteria also increases. $NH_3$ and $IC$ reduce due to the combined effect of these growth processes, while the growth of nitrifying bacteria also induces an increase in the concentration of $NO_3$. In the final phase of the cycle, a change in the slope of the $O_2$ concentration trend is observed, which increases again because $IC$ runs out and limits the activity of nitrifying bacteria and their $O_2$ consumption. The photosynthetic activity of microalgae and cyanobacteria involve a low release of $DOC$ which, however, is immediately degraded by heterotrophs. As can be seen from the concentrations at the end of the cycles, in this case nitrogen is not effectively removed from the wastewater, since about $50\%$ of $NH_3$ influent concentration is still present in the effluent and another significant nitrogen amount is found in the form of $NO_3$.

Finally, the trend of the concentration of soluble substrates within the SBR over time in the case of a municipal wastewater with presence of nitrate is shown in Fig. \ref{f5.4.2.3} (case 3: $S^{in}_{IC} = 180 \ g \  m^{-3}$, $S^{in}_{DOC} = 500 \ g \  m^{-3}$, $S^{in}_{NO_3} = 100 \ g \  m^{-3}$, $S^{in}_{NH_3} = 50 \ g \  m^{-3}$, $S^{in}_{O_2} = 0$). At the beginning of the cycle, there is a high concentration of $DOC$ and oxidized compounds ($O_2$ and $NO_3$) in the reactor and therefore there are optimal conditions for the growth of heterotrophs, first aerobic and then anoxic (when $O_2$ runs out). Consequently, the conversion of most of $DOC$ and $NH_3$ present is observed in the initial phase of the dark period. When concentrations of $O_2$ and $NO_3$ reduce, they limit heterotrophic kinetics and the consumption rate of $DOC$ and $NH_3$ slows down. However, as can be seen, $DOC$ has been almost completely removed from the system at the end of the dark period. When light conditions change, the photosynthetic activity of microalgae and cyanobacteria leads to the consumption of $NH_3$ and the increase of $O_2$ concentration. Since $DOC$ concentration is low, heterotrophic bacteria are not competitive and the oxygen produced is partially consumed by nitrifying bacteria, which convert the remaining $NH_3$ into $NO_3$. By observing the concentration of the soluble substrates at the end of the cycle, it is clear that the treatment cycle lead to the complete removal of $DOC$ and $NH_3$, while a $NO_3$ concentration of about $20 \ g \  m^{-3}$ is found in the effluent.

Fig. \ref{f5.4.2.4} shows the biomass distribution along the granule radius at $50$ days, in the four cases described until now. In the case of a typical municipal wastewater (top left) and in the case of a higher strength carbon wastewater (top right) the microbial distribution is very similar. To explain this result it is necessary to note that the two influent wastewaters differ only in the $DOC$ load. In the latter case, the growth kinetics of heterotrophic bacteria slow down during the light period, due to the depletion of $NH_3$ (as can be seen in Fig. \ref{f5.4.2.1}). Consequently, most of additional amount of $DOC$ present in this case is not consumed due to the lack of $NH_3$ and does not lead to significant differences in the growth and distribution of biomass within the photogranule reported in the case of a typical municipal wastewater. In the case of an ammonia wastewater (bottom left), results are totally different. $DOC$ is not present in the influent, hence, the only available carbon source for heterotrophs derives from the photoautotrophic release during the light period. However, such release is limited and just allows the growth of small amounts of heterotrophic biomass. Conversely, as there is no spatial competition with heterotrophs, high fractions of both cyanobacteria and microalgae are observed, and a small fraction of nitrifying bacteria, which is almost zero in the other cases, is also visible. 

 Anyway, due to the shortage of $DOC$ and the low heterotrophic contribution to the growth process, granules which develop in this case are reasonably smaller than in the other cases. Lastly, in the case of a municipal wastewater with presence of nitrate (bottom right), granules are populated by high fractions of heterotrophs, higher than previous cases, while microalgae are almost absent and cyanobacteria fraction is low and limited to the outermost layers. This is due to the high concentration of $NO_3$ present in the influent which, in anoxic conditions, replaces $O_2$ as the oxidized substrate for the metabolic activity of anoxic heterotrophs. Consequently, the growth of heterotrophs is not related exclusively to the production of $O_2$ by microalgae and cyanobacteria, but also occurs when there is no $O_2$, such as during the dark period (Fig. \ref{f5.4.2.3}). The sum of aerobic and anoxic growth processes leads to larger fractions of heterotrophs than the previous cases. In conclusion, in the first three cases, the formation and evolution of photogranules is strongly related to the photosynthetic activity, while in the last case the denitrification processes become dominant and take over the photoautotrophic processes.
 
\subsection{Study 3 - Effects of light conditions on the process} \label{n5.3.3}
\
Light governs the metabolic activity of phototrophic organisms and the photogranules evolution, and therefore represents a key factor in the OPG-based system. Moreover, light conditions affect the relative abundance of cyanobacteria and microalgae within the granule and also their ratio. As reported in Section \ref{n5.1}, microalgae and cyanobacteria have distinct modes of light harvesting and utilization which lead to different ecological niches. Specifically, under high light intensities the growth of microalgae is favored over cyanobacteria, which, however, are able to grow even under unfavorable light conditions. This numerical study is aimed to prove that the model can correctly describe the effects of light on microalgae and cyanobacteria growth and on the performances of the system. For this purpose, various simulations have been carried out by varying the incident light intensity $I_0$. The same influent composition of the first study is considered: $S^{in}_{IC} = 180 \ g \  m^{-3}$, $S^{in}_{DOC} = 500 \ g \  m^{-3}$, $S^{in}_{NH_3} = 50 \ g \  m^{-3}$, $S^{in}_{NO_3} = S^{in}_{O_2} = 0$.

\begin{figure*}
 \fbox{\includegraphics[width=1\textwidth, keepaspectratio]{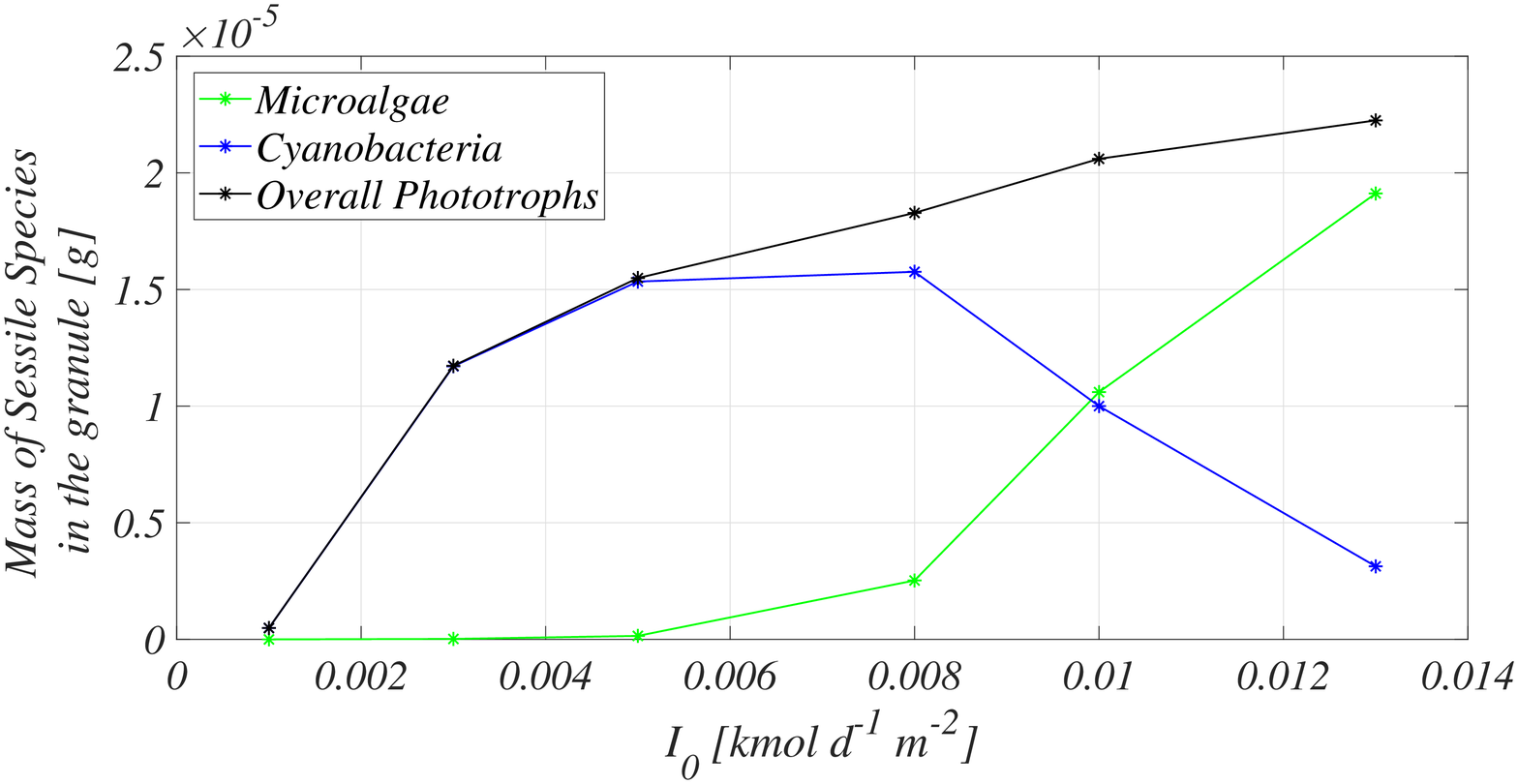}}   
 \caption{Study 3 - Mass of overall phototrophs, cyanobacteria and microalgae within the granule at $T = 50 \ d$ to vary the incident light intensity $I_0$. Wastewater influent composition: $S^{in}_{IC} = 180 \ g \  m^{-3}$ (inorganic carbon), $S^{in}_{DOC} = 500 \ g \  m^{-3}$ (organic carbon), $S^{in}_{NH_3} = 50 \ g \  m^{-3}$ (ammonia), $S^{in}_{NO_3} = 0 \ g \  m^{-3}$ (nitrate), $S^{in}_{O_2} = 0 \ g \  m^{-3}$ (oxygen).}
          \label{f5.4.3.1} 
 \end{figure*}

\begin{figure*}
 \fbox{\includegraphics[width=1\textwidth, keepaspectratio]{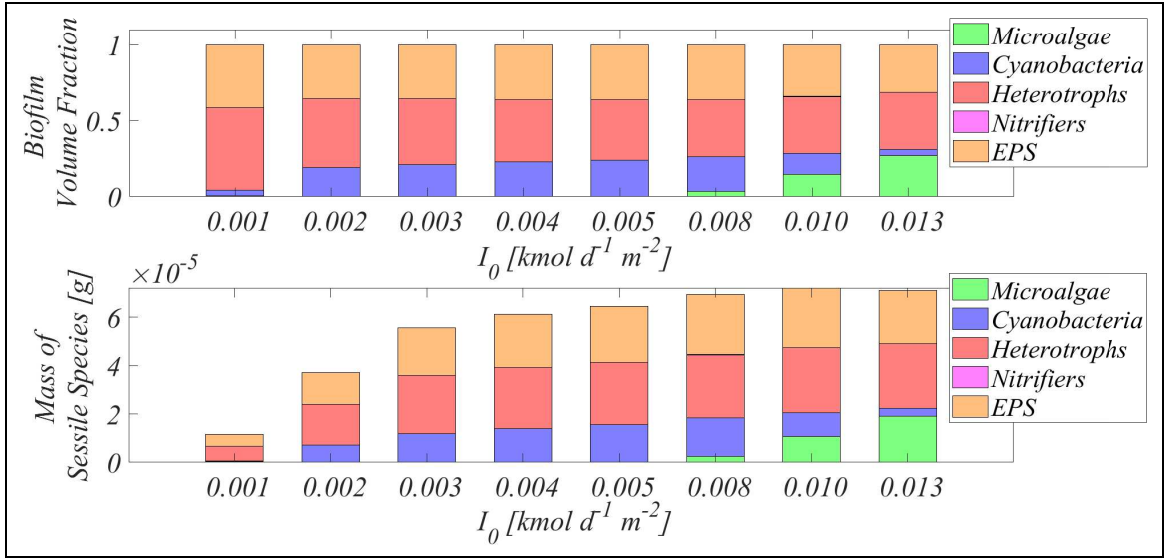}}   
 \caption{Study 3 - Relative abundances (top) and mass (bottom) of microbial species within the granule at $T = 50 \ d$ under different light conditions. Wastewater influent composition: $S^{in}_{IC} = 180 \ g \  m^{-3}$ (inorganic carbon), $S^{in}_{DOC} = 500 \ g \  m^{-3}$ (organic carbon), $S^{in}_{NH_3} = 50 \ g \  m^{-3}$ (ammonia), $S^{in}_{NO_3} = 0 \ g \  m^{-3}$ (nitrate), $S^{in}_{O_2} = 0 \ g \  m^{-3}$ (oxygen).}
          \label{f5.4.3.2} 
 \end{figure*}

Fig. \ref{f5.4.3.1} and Fig. \ref{f5.4.3.2} show results in terms of microbial composition within the photogranule under different light conditions. Specifically, Fig. \ref{f5.4.3.1} shows the overall mass of phototrophs and the mass of cyanobacteria and microalgae, while Fig. \ref{f5.4.3.2} shows the relative abundance (top) and mass (bottom) of sessile microbial species within the granule. The numerical results of both figures refer to $T=50 \ d$. The overall phototrophic mass (sum of cyanobacteria and microalgae) increases with $I_0$ (Fig. \ref{f5.4.3.1}), while the masses of cyanobacteria and microalgae are highly variable. The mass of microalgae increases with $I_0$ because their light dependency coefficient is directly proportional to $I_0$ in the range of values investigated (photoinhibition phenomena do not occur at these $I_0$ values). However, the mass of microalgae is very low up to $I_0 = 0.008 \ kmol \ m^{-2} \ d^{-1}$ and limited by the competition with cyanobacteria, which have the ability to adapt to not optimal light conditions. The mass of cyanobacteria reaches the maximum value for $I_0 = 0.008 \ kmol \ m^{-2} \ d^{-1}$ and decreases for higher values due to two reasons: optimal light conditions for cyanobacteria are supposed to be lower, hence, photoinhibition phenomena occur and limit the cyanobacterial growth; in addition, at these high light intensities, cyanobacteria suffer the competition with microalgae, which find optimal conditions to grow. Finally, in the case of $I_0 = 0.001 \ kmol \ m^{-2} \ d^{-1}$, light conditions are too poor even for cyanobacterial metabolic activity and small amounts of phototrophic mass and small granule dimension are observed. Indeed, due to a very low incident light intensity, the photosynthetic activity of microalgae and cyanobacteria is limited and leads to low $O_2$ productions. This also influences the growth of the other active biomasses. The result is the formation of small granules. Finally, in Fig. \ref{f5.4.3.2}, it can be noted that high fractions of heterotrophic bacteria and EPS are found in the photogranule for all $I_0$ values, while the fraction of nitrifying bacteria is not visible in any case.

       \begin{figure*}
 \fbox{\includegraphics[width=1\textwidth, keepaspectratio]{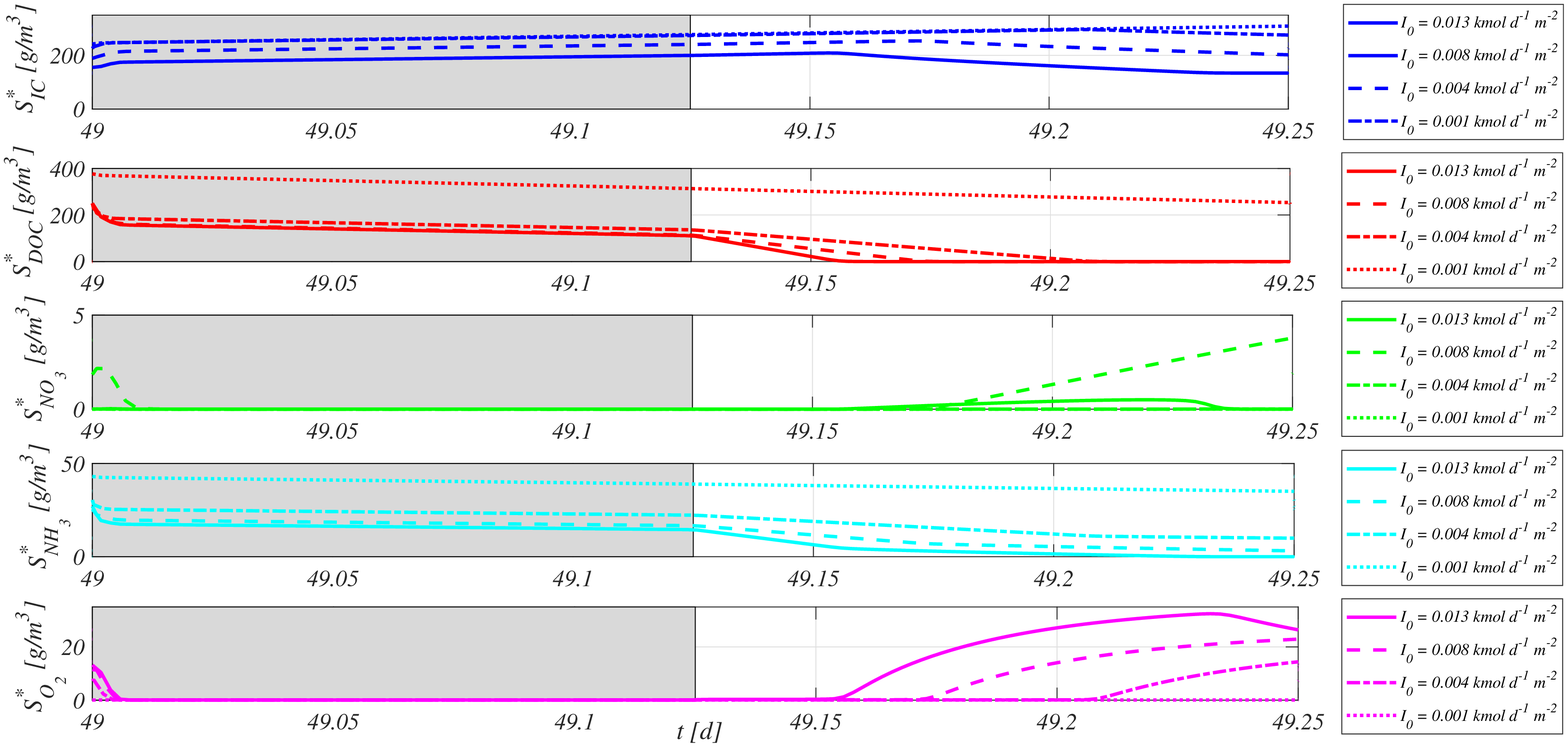}}   
 \caption{Study 3 - Evolution of soluble substrates concentration within the reactor, from $T=49 \ d$ to $T=50 \ d$ (four consecutive six-hours treatment cycles), under four different light conditions. Wastewater influent composition: $S^{in}_{IC} = 180 \ g \  m^{-3}$ (inorganic carbon), $S^{in}_{DOC} = 500 \ g \  m^{-3}$ (organic carbon), $S^{in}_{NH_3} = 100 \ g \  m^{-3}$ (ammonia), $S^{in}_{NO_3} = 0 \ g \  m^{-3}$ (nitrate), $S^{in}_{O_2} = 0 \ g \  m^{-3}$ (oxygen). Grey portions indicate the dark phases, white portions indicate the light phases.}
          \label{f5.4.3.3} 
 \end{figure*}

Fig. \ref{f5.4.3.3} reports the trend of the substrates concentration in the reactor under four different light conditions ($I_0 = 0.001-0.004-0.008-0.013 \ kmol \ m^{-2} \ d^{-1}$). The period shown is one complete cycle (six hours), starting from day $49$. Results of the four simulations differ in the light conditions which are provided in the second phase of the cycle (light period), while there are no differences during the dark phase, when a null value of $I_0$ is set for all cases. For this reason, trends of the substrates concentration are very similar during the dark period and differ mainly in the values at the beginning of the cycle, while significant differences are evident during the light phase. As $I_0$ increases going from $0.001$ to $0.013 \ kmol \ m^{-2} \ d^{-1}$, the rate of photosynthetic processes increases, hence, more $O_2$ is produced and more $IC$ is consumed. The greater production of $O_2$ favors the microbial activity of heterotrophic bacteria and leads to the faster consumption of $DOC$ and $NH_3$. In the case of $I_0 = 0.001 \ kmol \ m^{-2} \ d^{-1}$, the treatment cycle does not take place efficiently and large quantities of $DOC$ and $NH_3$ are found in the effluent, while in the other three cases, $DOC$ is totally removed and $NH_3$ keeps very low. The trend of $NO_3$ requires some observations. Although in small quantities, $NO_3$ is produced in the case of $I_0 = 0.008 \ kmol \ m^{-2} \ d^{-1}$, while in the other cases $NO_3$ is nearly zero throughout the cycle. When light conditions are poor ($I_0 = 0.001-0.004 \ kmol \ m^{-2} \ d^{-1}$), $DOC$ and $NH_3$ are present in the reactor throughout the cycle or almost, therefore the activity of heterotrophic bacteria strongly limits the nitrifying bacteria growth, and $NO_3$ production is not observed. On the other hand, when light conditions are high ($I_0 = 0.013 \ kmol \ m^{-2} \ d^{-1}$), the consumption of $DOC$ and $NH_3$ is faster and is completed before the end of the cycle. Again, nitrifying bacteria are unable to grow and to produce $NO_3$ because they are limited by the absence of $NH_3$. In conclusion, $I_0 = 0.008 \ kmol \ m^{-2} \ d^{-1}$ is the only case in which $DOC$ runs out and not $NH_3$, and optimal conditions for the growth of nitrifying bacteria and the production of small amounts of $NO_3$ occur within the system. In the case of $I_0 = 0.013 \ kmol \ m^{-2} \ d^{-1}$, a reduction of $O_2$ concentration is observed in the final part of the light phase. This occurs when both $NH_4$ and $NO_3$ are depleted and the photoautotrophic activity is limited by the lack of nutrients. Thus, the $O_2$ production reduces and the $O_2$ concentration decreases due to the effect of the gas transfer from bulk liquid to the atmosphere.

\section{Discussion and conclusions} \label{n5.4}
\
In this section the results shown above are discussed, with the aim of outlining the main aspects of OPGs and the OPGs-based system and drawing general conclusions from the numerical studies carried out.

This model allows to simulate the formation and evolution of oxygenic photogranules within an SBR reactor and describe the microbial composition of such granular biofilms. In the case of a OPG-based system fed with a typical municipal wastewater, granules are populated by high amounts of cyanobacteria. Their fraction is highest in the most superficial layers, and reduces towards the center of the granule, where disadvantageous light conditions occur due to attenuation phenomena. However, their presence is also found in the middle layers of the granule because, as it is known, cyanobacteria have the ability to proliferate even under poor or non-optimal light conditions \cite{tilzer1987light}. On the other hand, adaptation of microalgae to non-optimal light conditions is lower \cite{tilzer1987light}, and consequently their fraction is smaller than cyanobacteria and confined to the outermost layers of the granule. These results reflect what has been observed in \cite{abouhend2018oxygenic,abouhend2019growth} under the same light and operating conditions used in this numerical study. Moreover, large amounts of heterotrophs are observed while nitrifying bacteria are barely visible. This is due to the higher maximum growth rate of heterotrophs which, in presence of $DOC$, are more competitive from nitrifiers in the use of $O_2$ produced during the photosynthetic activity. Since the metabolic activity of nitrifying bacteria is negligible, concentration of $NO_3$ in the system is nearly zero throughout the cycle. Finally, a significant fraction of EPS is found throughout the granule, produced by all active microbial species and representing the "glue" that provides solidity and robustness to the aggregate. In conclusion, as suggested in \cite{milferstedt2017importance,ansari2019effects}, the model confirm that the treatment process is governed by the symbiotic interaction between heterotrophs and phototrophs: phototrophs produce $O_2$ necessary for heterotrophs, which degrade the polluting compounds from the wastewater. The metabolic activity of heterotrophs is especially relevant in the light phase, when the photosynthetic activity of phototrophs guarantees high concentrations of $O_2$ in the system. Conversely, during the dark phase, all microbial growth processes slow down. The $DOC/NH_3$ ratio in the influent is the key parameter to achieve the efficient treatment of the wastewater and simultaneously remove $DOC$ and $NH_3$ in the desired quantities. If such ratio is too high (Study 2, case 1), $NH_3$ runs out during the cycle (when $DOC$ is still present), becomes limiting for the kinetics of heterotrophs and phototrophs, and $DOC$ removal does not take place efficiently. If such ratio is too low (Study 2, case 2), excess $NH_3$ is used in the metabolic pathway of nitrifying bacteria, and converted into $NO_3$, and a nitrogen amount remains in the effluent in oxidized form.

The granule microbial composition described in the case of a typical municipal wastewater can totally change as the composition of the influent changes. For example, in the case of an ammonia rich wastewater without $DOC$ (Study 2, case 3), the growth of heterotrophs is severely limited and is exclusively supported by the low phototrophic $DOC$ release during the photosynthetic activity. Consequently, the heterotrophic fraction is very small and, on the other hand, higher fractions of cyanobacteria and microalgae are observed. Without heterotrophic competition, nitrifying bacteria find optimal conditions to grow and this time they are visible in the granule, albeit in very low fractions due to their low growth rate. An intense metabolic activity of nitrifying bacteria does not allow the effective removal of nitrogen from the wastewater, because it leads to the production of high $NO_3$ amounts  which are found in the effluent.

In the case of a municipal wastewater with high presence of $NO_3$ (Study 2, case 3), heterotrophic growth is supported especially during the dark period. Indeed, in absence of light and photosynthetic activity, $O_2$ runs out quickly and the subsequent anoxic conditions favor the growth of denitrifying heterotrophs, through the oxidation of high $NO_3$ amounts present within the reactor. The result is a photogranule constituted by high fractions of heterotrophs which take over cyanobacteria and microalgae, present in extremely small fractions and limited to the most superficial layers. Therefore, in this case the treatment cycle is mostly governed by denitrification processes and not by aerobic processes. It should be emphasized that the latter is a theoretical result suggested by the model, that should be supported by experimental evidence. Indeed, as it is known, cyanobacteria play a key role in the formation of photogranules and preservation of the spherical structure, but in such case their modest presence may not guarantee the development and the preservation of the granular structure.

Furthermore, the model outlines the key role of light conditions on the treatment process occurring in OPG-based systems, by analyzing their effects on the evolution of the granules, on the growth of the overall phototrophic biomass, on the relative abundance of cyanobacteria and microalgae and on the performances of the system. Poor light conditions severely affect the metabolic activity of all active biomasses: they limit the photosynthetic activity and the phototrophic growth, and lead to low $O_2$ productions, not sufficient to support the growth of heterotrophic and nitrifying bacteria. Therefore, granules formed have small dimension and the global biomass present in the system only partially remove the polluting compounds from the influent wastewater. Under better light conditions, the photosynthetic activity takes place more effectively and leads to the optimal growth of phototrophs and high productions of $O_2$, necessary for the metabolism of aerobic species. All this leads to the development of larger granules, mainly governed by heterotrophs, cyanobacteria and microalgae. In particular, due to their different light harvesting and utilization, the relative abundance of cyanobacteria and microalgae within the granule is highly variable to vary light conditions. Appreciable light conditions ($I_0 = 0.008 \ kmol \ m^{-2} \ d^{-1}$) guarantee the proliferation of cyanobacteria, which have a lower maximum growth rate than microalgae but greater abilities to adapt to adverse light conditions. Instead, very high incident light intensities favor the photosynthetic activity of microalgae, which take over the cyanobacteria. The described result confirms that the model is able to reproduce the behaviour of cyanobacteria and microalgae reported by several experimental studies \cite{abouhend2018oxygenic,abouhend2019growth}.

A significant drawback of the model is that it is not suitable to describe the uncertainty of the granulation phenomenon. As reported by \cite{hann2018factors,downes2019success}, photogranulation is an extremely complex process that is not always successful. Sometimes, it fails, without leading to the genesis of the granule, and the suspended inoculum remains in its original form. In this context, as specified in Section \ref{n5.1}, chances of success are deeply affected by the cyanobacteria concentration: high concentration of cyanobacteria in the system considerably increases the chances of photogranule formation. The model presented in this work describes the granulation process using a deterministic approach, through an attachment flux dependent on the concentration of cyanobacteria in suspended form. Such concentration is the discriminating factor: assuming a non-zero initial value leads to a non-zero attachment flux and the granule is formed, while assuming a zero value has the effect that all biomass present in the system remains in suspended form. In conclusion, in this model the initial concentration of cyanobacteria in suspended form determines the occurrence of the photogranulation and quantitatively affects the growth rate of the granule, but cannot affect the probability that the granule is formed.

In summary, it is possible to draw a number of conclusions from the numerical results presented:

\begin{itemize}
 \item The OPG-based system represents an interesting technology for the wastewater treatment, because it allows to remove aerobically the polluting compounds thanks to the production of $O_2$ by phototrophic microorganisms, without the use of external aeration sources.
  \item In this system, the removal of polluting compounds from the wastewater is based on the mechanisms of symbiotic interaction established between different microbial species. Among these, the most relevant is certainly the photoautotrophic production of $O_2$ by cyanobacteria and microalgae, necessary for the metabolic activity of heterotrophs.
 \item The influent composition and the influent carbon/nitrogen ratio strongly affect the efficiency of the system. As this ratio varies, carbon or nitrogen can become limiting for microbial kinetics and compromise the purifying efficiency of the photogranules.
 \item The influent composition radically influences the dynamics and the development of photogranules. Different concentrations of $DOC$, $NH_4$ and $NO_3$ lead to different microbial compositions within the photogranules. Phototrophic microorganisms (cyanobacteria and microalgae) and heterotrophic bacteria are the most present species in photogranules. However, nitrifying bacteria can proliferate in the case of poor DOC wastewaters.
 \item Main biological processes involved in the OPG-based system are photosynthetic processes and aerobic growth processes. However, high concentrations of $NO_3$ in the influent can lead to intense denitrification processes during the dark phase.
 \item Light conditions play a key role in the process by regulating the photosynthetic activity of phototrophic microorganisms, the production of $O_2$ and, indirectly, the growth of other microbial species. Consequently, light conditions significantly affects the dimension of photogranules and the performances of the system.
 \end{itemize}

 \section{Appendix A} \label{n5.5}
 \
All reaction terms of the model are written below. The kinetic expressions of the biological processes (in the biofilm $\nu$ and in the bulk liquid $\nu^*$) and the stoichiometric coefficients (in the biofilm $\alpha$ and in the bulk liquid $\alpha^*$) are reported in Tables \ref{t5.6.1}, \ref{t5.6.2}, \ref{t5.6.3} and \ref{t5.6.4}.

  \begin{equation}                                        \label{5.6.1}
r_{M,A} =  \alpha_{A,1} \ \nu_{A,1} + \alpha_{A,2} \ \nu_{A,2} + \alpha_{A,3} \ \nu_{A,3} - k_{d,A} \ f_{A},
 \end{equation}
 
  \begin{equation}                                        \label{5.6.2}
r_{M,C} =  \alpha_{C,1} \ \nu_{C,1} + \alpha_{C,2} \nu_{C,2} + \alpha_{C,3} \nu_{C,3} - k_{d,C} \ f_{C},
 \end{equation}
 
   \begin{equation}                                        \label{5.6.3}
r_{M,H} =  \alpha_{H,1} \ \nu_{H,1} + \alpha_{H,2} \ \nu_{H,2} - k_{d,H} \ f_{H},
 \end{equation}

  \begin{equation}                                        \label{5.6.4}
r_{M,N} =  \alpha_{N,1} \ \nu_{N} - k_{d,N} \ f_{N},
 \end{equation}

\[
r_{M,EPS} = \alpha_{EPS,1}  \ \nu_{A,1} + \alpha_{EPS,2}  \ \nu_{A,2} + \alpha_{EPS,3}  \ \nu_{C,1} + \alpha_{EPS,4} \ \nu_{C,2} + \alpha_{EPS,5}  \ \nu_{H,1} + 
\]
  \begin{equation}                                         \label{5.6.5}
 + \alpha_{EPS,6}  \ \nu_{H,2} + \alpha_{EPS,7}  \ \nu_{N},
 \end{equation}

  \begin{equation}                                        \label{5.6.6}
r_{M,I} =  k_{d,A} \ f_{A}  + k_{d,C} \ f_{C} + k_{d,H} \ f_{H} + k_{d,N} \ f_{N},
 \end{equation}

\[
r_{S,IC} = \alpha_{IC,1}  \ \nu_{A,1} \rho + \alpha_{IC,2}  \ \nu_{A,2} \rho + \alpha_{IC,3} \ \nu_{A,3} \rho   + \alpha_{IC,4}  \ \nu_{C,1} \rho +
\]
  \begin{equation}                                        \label{5.6.7}
+\alpha_{IC,5} \ \nu_{C,2} \rho + \alpha_{IC,6} \ \nu_{C,3} \rho+ \alpha_{IC,7}  \ \nu_{H,1} \rho + \alpha_{IC,8}  \ \nu_{H,2} \rho + \alpha_{IC,9}  \ \nu_{N} \rho,
\end{equation}

\[
r_{S,DOC} = \alpha_{DOC,1}  \ \nu_{A,1} \rho + \alpha_{DOC,2}  \ \nu_{A,2} \rho + \alpha_{DOC,3} \ \nu_{A,3} \rho + \alpha_{DOC,4}  \ \nu_{C,1} \rho +
\]
  \begin{equation}                                        \label{5.6.8}
 + \alpha_{DOC,5} \ \nu_{C,2} \rho + \alpha_{DOC,6} \ \nu_{C,3} \rho + \alpha_{DOC,7}  \ \nu_{H,1} \rho + \alpha_{DOC,8}  \ \nu_{H,2} \rho,
\end{equation}

  \begin{equation}                                        \label{5.6.9}
r_{S,NO3} = \alpha_{NO3,1}  \ \nu_{A,2} \rho  + \alpha_{NO3,2} \ \nu_{C,2} \rho + \alpha_{NO3,3}  \ \nu_{H,2} \rho + \alpha_{NO3,4}  \ \nu_{N} \rho,
\end{equation}

\[
r_{S,NH3} = \alpha_{NH3,1}  \ \nu_{A,1} \rho + \alpha_{NH3,2} \ \nu_{A,3} \rho  + \alpha_{NH3,3}  \ \nu_{C,1} \rho + \alpha_{NH3,4} \ \nu_{C,3} \rho +
\]
  \begin{equation}                                        \label{5.6.10}
+ \alpha_{NH3,5}  \ \nu_{H,1} \rho + \alpha_{NH3,6}  \ \nu_{H,2} \rho + \alpha_{NH3,7}  \ \nu_{N} \rho,
\end{equation}

\[
r_{S,O2} = \alpha_{O2,1}  \ \nu_{A,1} \rho + \alpha_{O2,2}  \ \nu_{A,2} \rho + \alpha_{O2,3} \ \nu_{A,3} \rho + \alpha_{O2,4}  \ \nu_{C,1} \rho +
\]
  \begin{equation}                                        \label{5.6.11}
+ \alpha_{O2,5} \ \nu_{C,2} \rho + \alpha_{O2,6} \ \nu_{C,3} \rho + \alpha_{O2,7}  \ \nu_{H,1} \rho + \alpha_{O2,8}  \ \nu_{N} \rho,
\end{equation}

  \begin{equation}                                        \label{5.6.12}
r^*_{\psi,A} =  \alpha^*_{A,1} \ \nu^*_{A,1} + \alpha^*_{A,2} \nu^*_{A,2} + \alpha^*_{A,3} \nu^*_{A,3} - k_{d,A} \ \psi^*_{A},
 \end{equation}
 
   \begin{equation}                                        \label{5.6.13}
r^*_{\psi,C} =  \alpha^*_{C,1} \nu^*_{C,1} + \alpha^*_{C,2} \nu^*_{C,2} + \alpha^*_{C,3} \nu^*_{C,3} - k_{d,C} \ \psi^*_{C},
 \end{equation}

   \begin{equation}                                        \label{5.6.14}
r^*_{\psi,H} =  \alpha^*_{H,1} \ \nu^*_{H,1} + \alpha^*_{H,2} \nu^*_{H,2} - k_{d,H} \ \psi^*_{H},
 \end{equation}

  \begin{equation}                                        \label{5.6.15}
r^*_{\psi,N} =  \alpha^*_{N,1} \nu^*_{N} - k_{d,N} \ \psi^*_{N},
 \end{equation}

\[
r^*_{S,IC} = \alpha^*_{IC,1} \nu^*_{A,1} + \alpha^*_{IC,2} \ \nu^*_{A,2} + \alpha^*_{IC,3} \ \nu^*_{A,3} + \alpha^*_{IC,4} \ \nu^*_{C,1} + \alpha^*_{IC,5} \ \nu^*_{C,2} +
\]
  \begin{equation}                                        \label{5.6.16}
+ \alpha^*_{IC,6} \ \nu^*_{C,3} + \alpha^*_{IC,7} \ \nu^*_{H,1} + \alpha^*_{IC,8} \ \nu^*_{H,2} + \alpha^*_{IC,9} \ \nu^*_{N},
\end{equation}

\[
r^*_{S,DOC} = \alpha^*_{DOC,1} \ \nu^*_{A,1} + \alpha^*_{DOC,2} \ \nu^*_{A,2} + \alpha^*_{DOC,3} \ \nu^*_{A,3} + \alpha^*_{DOC,4} \ \nu^*_{C,1} +
\]
  \begin{equation}                                        \label{5.6.17}
 + \alpha^*_{DOC,5} \ \nu^*_{C,2} + \alpha^*_{DOC,6} \ \nu^*_{C,3} + \alpha^*_{DOC,7} \ \nu^*_{H,1} + \alpha^*_{DOC,8} \ \nu^*_{H,2},
\end{equation}

  \begin{equation}                                        \label{5.6.18}
r^*_{S,NO3} = \alpha^*_{NO3,1} \ \nu^*_{A,2}  + \alpha^*_{NO3,2} \ \nu^*_{C,2} + \alpha^*_{NO3,3} \ \nu^*_{H,2} + \alpha^*_{NO3,4} \ \nu^*_{N},
\end{equation}

\[
r^*_{S,NH3} = \alpha^*_{NH3,1} \ \nu^*_{A,1} + \alpha^*_{NH3,2} \ \nu^*_{A,3}  + \alpha^*_{NH3,3} \ \nu^*_{C,1} + \alpha^*_{NH3,4} \ \nu^*_{C,3} +
\]
  \begin{equation}                                        \label{5.6.19}
 + \alpha^*_{NH3,5} \ \nu^*_{H,1} + \alpha^*_{NH3,6} \ \nu^*_{H,2} + \alpha^*_{NH3,7} \ \nu^*_{N},
\end{equation}

\[
r^*_{S,O2} = \alpha^*_{O2,1} \ \nu^*_{A,1} + \alpha^*_{O2,2} \ \nu^*_{A,2} + \alpha^*_{O2,3} \ \nu^*_{A,3}  + \alpha^*_{O2,4} \ \nu^*_{C,1} +
\]
  \begin{equation}                                        \label{5.6.20}
 + \alpha^*_{O2,5} \ \nu^*_{C,2} + \alpha^*_{O2,6} \ \nu^*_{C,3} + \alpha^*_{O2,7} \ \nu^*_{H,1} + \alpha^*_{O2,8} \ \nu^*_{N} + k_{La} \ (S_{O2,sat}-S^*_{O_2}).
\end{equation}

  \begin{table*}[p]
\begin{footnotesize}
\begin{spacing}{2}
 \begin{center}
 \medskip
\rotatebox{90}{%
 \begin{tabular}{lc}
 \hline
{\textbf{Process}} & {\textbf{Expression}} 
 \\
 \hline
Photoautotrophic growth of microalgae on $NH_3$ & $\nu_{A,1}=q_{max,A} \frac{S_{IC}}{K_{A,IC}+S_{IC}} \frac{S_{NH3}}{K_{A,NH3}+S_{NH3}} \frac{K^{in}_{A,O2}}{K^{in}_{A,O2}+S_{O2}} (\frac{I}{I_{opt,A}})^{\eta_{A}} \ e^{(1-(\frac{I}{I_{opt,A}})^{\eta_{A}})} \ f_{A}$  \\
 
Photoautotrophic growth of microalgae on $NO_3$ & $\nu_{A,2}=q_{max,A} \frac{S_{IC}}{K_{A,IC}+S_{IC}} \frac{S_{NO3}}{K_{A,NO3}+S_{NO3}} \frac{K^{in}_{A,NH3}}{K^{in}_{A,NH3}+S_{NH3}} \frac{K^{in}_{A,O2}}{K^{in}_{A,O2}+S_{O2}} (\frac{I}{I_{opt,A}})^{\eta_{A}} \ e^{(1-(\frac{I}{I_{opt,A}})^{\eta_{A}})} \ f_{A}$  \\
  
Heterotrophic growth of microalgae & $\nu_{A,3}=0.1 \ q_{max,A} \frac{S_{DOC}}{K_{A,DOC}+S_{DOC}} \frac{S_{O2}}{K_{A,O2}+S_{O2}} \frac{K^{in}_{A,I}}{K^{in}_{A,I}+I} \ f_{A}$  \\
      
Photoautotrophic growth of cyanobacteria on $NH_3$ & $\nu_{C,1}=q_{max,C} \frac{S_{IC}}{K_{C,IC}+S_{IC}} \frac{S_{NH3}}{K_{C,NH3}+S_{NH3}} \frac{K^{in}_{C,O2}}{K^{in}_{C,O2}+S_{O2}} (\frac{I}{I_{opt,C}})^{\eta_C} \ e^{(1-(\frac{I}{I_{opt,C}})^{\eta_C)}} \ f_C $  \\
  
Photoautotrophic growth of cyanobacteria on $NO_3$ & $\nu_{C,2}=q_{max,C} \frac{S_{IC}}{K_{C,IC}+S_{IC}} \frac{S_{NO3}}{K_{C,NO3}+S_{NO3}} \frac{K^{in}_{C,NH3}}{K^{in}_{C,NH3}+S_{NH3}} \frac{K^{in}_{C,O2}}{K^{in}_{C,O2}+S_{O2}} (\frac{I}{I_{opt,C}})^{\eta_C} \ e^{(1-(\frac{I}{I_{opt,C}})^{\eta_C)}} \ f_C $  \\
  
Heterotrophic growth of cyanobacteria  & $\nu_{C,3}=0.1 \ q_{max,C} \frac{S_{DOC}}{K_{C,DOC}+S_{DOC}} \frac{S_{O2}}{K_{C,O2}+S_{O2}} \frac{K^{in}_{C,I}}{K^{in}_{C,I}+I} \ f_C$  \\
        
Aerobic growth of heterotrophic bacteria  & $\nu_{H,1}=\mu_{max,H} \frac{S_{DOC}}{K_{H,DOC}+S_{DOC}} \frac{S_{NH3}}{K_{H,NH3}+S_{NH3}} \frac{S_{O2}}{K_{H,O2}+S_{O2}}\ f_H$  \\
  
Anoxic growth of heterotrophic bacteria  & $\nu_{H,2}=\mu_{max,H} \frac{S_{DOC}}{K_{H,DOC}+S_{DOC}} \frac{S_{NO3}}{K_{H,NO3}+S_{NO3}} \frac{S_{NH3}}{K_{H,NH3}+S_{NH3}}  \frac{K_{H,O2}}{K_{H,O2}+S_{O2}} \ f_H$  \\
              
Growth of nitrifying bacteria  & $\nu_{N}=\mu_{max,N} \frac{S_{IC}}{K_{N,IC}+S_{IC}} \frac{S_{NH3}}{K_{H,NH3}+S_{NH3}} \frac{S_{O2}}{K_{H,O2}+S_{O2}} \ f_N$  \\
  \hline
  \multicolumn{2}{l}{where  $K^{in}_{A,O2}=K^{in}_{C,O2}=K^{in}_{O2,max} \ \frac{\frac{S_{IC}}{S_{O2}}}{\frac{S_{IC}}{S_{O2}}+K_{R_{CO2/O2}}}$} \\
  
 \end{tabular}%
 }
 \caption{Kinetic expressions of the growth processes within biofilm granules.} \label{t5.6.1}
 \end{center}
 \end{spacing}
 \end{footnotesize}
 \end{table*}

  \begin{table*}[p]
\begin{footnotesize}
\begin{spacing}{2}
 \begin{center}
 \medskip
\rotatebox{90}{%
 \begin{tabular}{lc}
 \hline
{\textbf{Process}} & {\textbf{Expression}} 
 \\
 \hline
Photoautotrophic growth of microalgae on $NH_3$ & $\nu^*_{A,1}=q_{max,A} \frac{S^*_{IC}}{K_{A,IC}+S^*_{IC}} \frac{S^*_{NH3}}{K_{A,NH3}+S^*_{NH3}} \frac{K^{in,*}_{A,O2}}{K^{in,*}_{A,O2}+S^*_{O2}} (\frac{I_0}{I_{opt,A}})^{\eta_{A}} \ e^{(1-(\frac{I_0}{I_{opt,A}})^{\eta_{A}})} \ \psi^*_{A}$  \\
  
Photoautotrophic growth of microalgae on $NO_3$ & $\nu^*_{A,2}=q_{max,A} \frac{S^*_{IC}}{K_{A,IC}+S^*_{IC}} \frac{S^*_{NO3}}{K_{A,NO3}+S^*_{NO3}} \frac{K^{in}_{A,NH3}}{K^{in}_{A,NH3}+S^*_{NH3}} \frac{K^{in,*}_{A,O2}}{K^{in,*}_{A,O2}+S^*_{O2}} (\frac{I_0}{I_{opt,A}})^{\eta_{A}} \ e^{(1-(\frac{I_0}{I_{opt,A}})^{\eta_{A}})} \ \psi^*_{A}$  \\
  
 Heterotrophic growth of microalgae & $\nu^*_{A,3} = 0.1 \ q_{max,A} \frac{S^*_{DOC}}{K_{A,DOC}+S^*_{DOC}} \frac{S^*_{O2}}{K_{A,O2}+S^*_{O2}} \frac{K^{in}_{A,I}}{K^{in}_{A,I}+I_0} \psi^*_{A}$  \\
      
Photoautotrophic growth of cyanobacteria on $NH_3$& $\nu^*_{C,1}=q_{max,C} \frac{S^*_{IC}}{K_{C,IC}+S^*_{IC}} \frac{S^*_{NH3}}{K_{C,NH3}+S^*_{NH3}} \frac{K^{in,*}_{C,O2}}{K^{in,*}_{C,O2}+S^*_{O2}} (\frac{I_0}{I_{opt,C}})^{\eta_C} \ e^{(1-\frac{I_0}{I_{opt,C}})^{\eta_C}} \ \psi^*_C$  \\
  
Photoautotrophic growth of cyanobacteria on $NO_3$ & $\nu^*_{C,2}=q_{max,C} \frac{S^*_{IC}}{K_{C,IC}+S^*_{IC}} \frac{S^*_{NO3}}{K_{C,NO3}+S^*_{NO3}} \frac{K^{in}_{C,NH3}}{K^{in}_{C,NH3}+S^*_{NH3}} \frac{K^{in,*}_{C,O2}}{K^{in,*}_{C,O2}+S^*_{O2}} (\frac{I_0}{I_{opt,C}})^{\eta_C} \ e^{(1-(\frac{I_0}{I_{opt,C}})^{\eta_C)}} \ \psi^*_C$  \\
  
Heterotrophic growth of cyanobacteria & $\nu^*_{C,3}=0.1 \ q_{max,C} \frac{S^*_{DOC}}{K_{C,DOC}+S^*_{DOC}} \frac{S^*_{O2}}{K_{C,O2}+S^*_{O2}} \frac{K^{in}_{C,I}}{K^{in}_{C,I}+I_0} \ \psi^*_C $  \\
     
Aerobic growth of heterotrophic bacteria & $\nu^*_{H,1}=\mu_{max,H} \frac{S^*_{DOC}}{K_{H,DOC}+S^*_{DOC}} \frac{S^*_{NH3}}{K_{H,NH3}+S^*_{NH3}} \frac{S^*_{O2}}{K_{H,O2}+S^*_{O2}}\ \psi^*_H$  \\
  
Anoxic growth of heterotrophic bacteria & $\nu^*_{H,2}=\mu_{max,H} \frac{S^*_{DOC}}{K_{H,DOC}+S^*_{DOC}} \frac{S^*_{NO3}}{K_{H,NO3}+S^*_{NO3}} \frac{S^*_{NH3}}{K_{H,NH3}+S^*_{NH3}}  \frac{K_{H,O2}}{K_{H,O2}+S^*_{O2}} \ \psi^*_H$  \\
              
Growth of nitrifying bacteria  & $\nu^*_{N}=\mu_{max,N} \frac{S^*_{IC}}{K_{N,IC}+S^*_{IC}} \frac{S^*_{NH3}}{K_{H,NH3}+S^*_{NH3}} \frac{S^*_{O2}}{K_{H,O2}+S^*_{O2}} \ \psi^*_N$  \\
  
 \hline
  \multicolumn{2}{l}{where $K^{in,*}_{A,O2}=K^{in,*}_{C,O2}=K^{in}_{O2,max} \ \frac{\frac{S^*_{IC}}{S^*_{O2}}}{\frac{S^*_{IC}}{S^*_{O2}}+K_{R_{CO2/O2}}}$}\\
  
 \end{tabular}%
 }
 \caption{Kinetic expressions of the growth processes within the bulk liquid.} \label{t5.6.2}
 \end{center}
 \end{spacing}
 \end{footnotesize}
 \end{table*}
 
   \begin{table}[ht]
\begin{footnotesize}
\begin{spacing}{1.6}
 \begin{center}
 \begin{tabular}{cccc}
 \hline
{\textbf{Stoich. coefficient}} & {\textbf{Expression}} & {\textbf{Stoich. coefficient}} & {\textbf{Expression}} 
 \\
 \hline
$\alpha_{A,1}$ & $\frac{32}{1+\phi_{EPS,A}+k_{DOC}}$  &$\alpha_{DOC,2}$ & $ \frac{32 \ k_{DOC}}{1.3409+\phi_{EPS,A}+k_{DOC}}$   \\
$\alpha_{A,2}$ & $\frac{32}{1.3409+\phi_{EPS,A}+k_{DOC}}$   &$\alpha_{DOC,3}$ & $ \frac{-32}{1-Y_{DOC}}$     \\   
$\alpha_{A,3}$ & $\frac{32 \ Y_{DOC}}{1-Y_{DOC}} $   &$\alpha_{DOC,4}$ & $\frac{32 \ k_{DOC}}{1.3409+\phi_{EPS,C}+k_{DOC}}$\\  
$\alpha_{C,1}$ & $\frac{32}{1+\phi_{EPS,C}+k_{DOC}}$ & $\alpha_{DOC,5}$ & $\frac{32 \ k_{DOC}}{1.3409+\phi_{EPS,C}+k_{DOC}}$  \\   
$\alpha_{C,2}$ & $\frac{32}{1.3409+\phi_{EPS,C}+k_{DOC}}$ &$\alpha_{DOC,6}$ & $\frac{-32}{1-Y_{DOC}}$     \\
$\alpha_{C,3}$ & $\frac{32 \ Y_{DOC}}{1-Y_{DOC}}$ &$\alpha_{DOC,7}$ & $  -\frac{1}{Y_H}$   \\
$\alpha_{H,1}$ & $1-k_{EPS,H}$  &$\alpha_{DOC,8}$ & $  -\frac{1}{Y_H}$     \\ 
$\alpha_{H,2}$ & $1-k_{EPS,H}$  &$\alpha_{NO3,1}$ & $- \frac{0.1704}{1.3409+\phi_{EPS,A}+k_{DOC}}$\\       
$\alpha_{N,1}$ & $1-k_{EPS,N}$ &$\alpha_{NO3,2}$ & $- \frac{0.1704}{1.3409+\phi_{EPS,C}+k_{DOC}}$\\        
$\alpha_{EPS,1}$ & $\frac{32 \ \phi_{EPS,A}}{1+\phi_{EPS,A}+k_{DOC}} $&$\alpha_{NO3,3}$ & $ -\frac{0.8}{32 \ Y_H}+0.02857$     \\    
$\alpha_{EPS,2}$ & $\frac{32 \ \phi_{EPS,A}}{1.3409+\phi_{EPS,A}+k_{DOC}}$&$\alpha_{NO3,4}$ & $+ \frac{1}{14 \ Y_N}$\\ 
$\alpha_{EPS,3}$ & $\frac{32 \ \phi_{EPS,C}}{1+\phi_{EPS,C}+k_{DOC}}$&$\alpha_{NH3,1}$ & $- \frac{0.1704}{1+\phi_{EPS,A}+k_{DOC}}$     \\
$\alpha_{EPS,4}$ & $\frac{32 \ \phi_{EPS,C}}{1.3409+\phi_{EPS,C}+k_{DOC}}$&$\alpha_{NH3,2}$ & $ - \frac{0.1704}{1-Y_{DOC}}$     \\
$\alpha_{EPS,5}$ & $k_{EPS,H}$&   $\alpha_{NH3,3}$ & $  - \frac{0.1704}{1+\phi_{EPS,C}+k_{DOC}}$\\  
$\alpha_{EPS,6}$ & $k_{EPS,H}$&    $\alpha_{NH3,4}$ & $ - \frac{0.1704}{1-Y_{DOC}}$     \\ 
$\alpha_{EPS,7}$ & $k_{EPS,N}$& $\alpha_{NH3,5}$ & $ - \frac{0.2}{33.6}$     \\
$\alpha_{IC,1}$ & $ - \frac{1.0025+\phi_{EPS,A}+k_{DOC}}{1+\phi_{EPS,A}+k_{DOC}}$&$\alpha_{NH3,6}$ & $ - \frac{0.2}{33.6}  $\\
$\alpha_{IC,2}$ & $ - \frac{1.0025+\phi_{EPS,A}+k_{DOC}}{1.3409+\phi_{EPS,A}+k_{DOC}}$& $\alpha_{NH3,7}$ & $  -\frac{1}{14 \ Y_N}-0.00593$ \\  
$\alpha_{IC,3}$ & $ \frac{1-1.0025 \ Y_{DOC}}{1-Y_{DOC}}$&  $\alpha_{O2,1}$ & $1$     \\ 
$\alpha_{IC,4}$ & $ - \frac{1.0025+\phi_{EPS,C}+k_{DOC}}{1+\phi_{EPS,C}+k_{DOC}}$&$\alpha_{O2,2}$ & $1$     \\
$\alpha_{IC,5}$ & $ - \frac{1.0025+\phi_{EPS,C}+k_{DOC}}{1.3409+\phi_{EPS,C}+k_{DOC}}$&$\alpha_{O2,3}$ & $- 1$     \\   
$\alpha_{IC,6}$ & $ \frac{1-1.0025 \ Y_{DOC}}{1-Y_{DOC}}$&$\alpha_{O2,4}$ & $1$\\
$\alpha_{IC,7}$ & $   \frac{1}{32 \ Y_H}-0.02976$&       $\alpha_{O2,5}$ & $1$\\ 
$\alpha_{IC,8}$ & $  \frac{1}{32 \ Y_H}-0.02976$&$\alpha_{O2,6}$ & $- 1$\\
$\alpha_{IC,9}$ & $   -\frac{1}{33.6}$&$\alpha_{O2,7}$ & $- \frac{1}{32 \ Y_H}+0.03125$\\ 
$\alpha_{DOC,1}$ & $  \frac{32 \ k_{DOC}}{A.3409+\phi_{EPS,A}+k_{DOC}}$&$\alpha_{O2,8}$ & $- \frac{1}{7 \ Y_N}+\frac{1}{32}$\\
 \hline 
     \end{tabular}
 \caption{Stoichiometric coefficients used in the model biological processes within biofilm granules} \label{t5.6.3}
 \end{center}
 \end{spacing}
 \end{footnotesize}
 \end{table}

\begin{table}[ht]
\begin{footnotesize}
\begin{spacing}{1.6}
 \begin{center}
 \begin{tabular}{cccc}
 \hline
{\textbf{Stoich. coefficient}} & {\textbf{Expression}}  &{\textbf{Stoich. coefficient}} & {\textbf{Expression}} 

 \\
 \hline
$\alpha^*_{A,1}$ & $\frac{32}{1+k_{DOC}}$&$\alpha^*_{DOC,6}$ & $ \frac{-32}{1-Y_{DOC}}$\\
$\alpha^*_{A,2}$ & $\frac{32}{1.3409+k_{DOC}}$ &$\alpha^*_{DOC,7}$ & $  -\frac{1}{Y_H}  $ \\
$\alpha^*_{A,3}$ & $\frac{32 \ Y_{DOC}}{1-Y_{DOC}} $ &$\alpha^*_{DOC,8}$ & $  -\frac{1}{Y_H} $\\
$\alpha^*_{C,1}$ & $\frac{32}{1+k_{DOC}}$ &$\alpha^*_{NO3,1}$ & $ - \frac{0.1704}{1.3409+k_{DOC}}$\\
$\alpha^*_{C,2}$ & $\frac{32}{1.3409+k_{DOC}}$&$\alpha^*_{NO3,2}$ & $ - \frac{0.1704}{1.3409+k_{DOC}}$\\
$\alpha^*_{C,3}$ & $\frac{32 \ Y_{DOC}}{1-Y_{DOC}}$ &$\alpha^*_{NO3,3}$ & $ - \frac{0.8}{32 \ Y_H}+0.02857$\\
$\alpha^*_{H,1}$ & $ 1$ &$\alpha^*_{NO3,4}$ & $ + \frac{1}{14 \ Y_N}$\\
$\alpha^*_{H,2}$ & $ 1$ &$\alpha^*_{NH3,1}$ & $  - \frac{0.1704}{1+k_{DOC}}$\\
$\alpha^*_{N,1}$ & $ 1$&$\alpha^*_{NH3,2}$ & $ - \frac{0.1704}{1-Y_{DOC}}$\\
$\alpha^*_{IC,1}$ & $  - \frac{1.0025+k_{DOC}}{1+k_{DOC}} $ &$\alpha^*_{NH3,3}$ & $  - \frac{0.1704}{1+k_{DOC}}$\\
$\alpha^*_{IC,2}$ & $ - \frac{1.0025+k_{DOC}}{1.3409+k_{DOC}}$ &$\alpha^*_{NH3,4}$ & $ - \frac{0.1704}{1-Y_{DOC}}$\\
$\alpha^*_{IC,3}$ & $ \frac{1-1.0025 \ Y_{DOC}}{1-Y_{DOC}}$ &$\alpha^*_{NH3,5}$ & $ - \frac{0.2}{33.6}$\\
$\alpha^*_{IC,4}$ & $ - \frac{1.0025+k_{DOC}}{1+k_{DOC}}$ &$\alpha^*_{NH3,6}$ & $ - \frac{0.2}{33.6}  $\\
$\alpha^*_{IC,5}$ & $ - \frac{1.0025+k_{DOC}}{1.3409+k_{DOC}}$ &$\alpha^*_{NH3,7}$ & $  -\frac{1}{14 \ Y_N}-0.00593$\\
$\alpha^*_{IC,6}$ & $ \frac{1-1.0025 \ Y_{DOC}}{1-Y_{DOC}}$ &$\alpha^*_{O2,1}$ & $  1$\\
$\alpha^*_{IC,7}$ & $   \frac{1}{32 \ Y_H}-0.02976$ &$\alpha^*_{O2,2}$ & $ 1$\\
$\alpha^*_{IC,8}$ & $   \frac{1}{32 \ Y_H}-0.02976$ &$\alpha^*_{O2,3}$ & $ - 1$\\
$\alpha^*_{IC,9}$ & $   -\frac{1}{33.6}$ &$\alpha^*_{O2,4}$ & $  1$\\
$\alpha^*_{DOC,1}$ & $  \frac{32 \ k_{DOC}}{1.3409+k_{DOC}}$ &$\alpha^*_{O2,5}$ & $ 1$\\
$\alpha^*_{DOC,2}$ & $ \frac{32 \ k_{DOC}}{1.3409+k_{DOC}}$ &$\alpha^*_{O2,6}$ & $ - 1$\\
$\alpha^*_{DOC,3}$ & $ \frac{-32}{1-Y_{DOC}}$ &$\alpha^*_{O2,7}$ & $ - \frac{1}{32 \ Y_H}+0.03125 $\\
$\alpha^*_{DOC,4}$ & $  \frac{32 \ k_{DOC}}{1.3409+k_{DOC}}$&$\alpha^*_{O2,8}$ & $ - \frac{1}{7 \ Y_N}+\frac{1}{32}$\\
$\alpha^*_{DOC,5}$ & $ \frac{32 \ k_{DOC}}{1.3409+k_{DOC}}$&
\\
 \hline 
     \end{tabular}
 \caption{Stoichiometric coefficients used in the model for biological processes within the bulk liquid} \label{t5.6.4}
 \end{center}
 \end{spacing}
 \end{footnotesize}
 \end{table}

\clearpage
\bibliographystyle{unsrt}      
\bibliography{Tenore_photogranule}

\end{document}